\documentclass[twocolumn,floatfix,amsmath,amssymb,superscriptaddress,nofootinbib,longbibliography]{revtex4-2}

\usepackage{lipsum}
\usepackage{verbatim}
\usepackage[colorlinks=true,linkcolor=blue,citecolor=blue]{hyperref}
\usepackage[utf8]{inputenc}
\usepackage[T1]{fontenc}

\usepackage{bm}
\usepackage{array}
\usepackage{mathtools}
\usepackage[mathscr]{euscript}

\usepackage{physics}
\usepackage{algorithmic}
\usepackage{algorithm}
\usepackage{siunitx}

\usepackage{graphicx}
\usepackage{tikz}
\usepackage{pgfplots}
\usepackage[caption=false]{subfig}
\usepackage{quantikz}
\usetikzlibrary{quantikz}
\usetikzlibrary{fadings}

\usepackage{tabularx}
\usepackage{booktabs}

\DeclarePairedDelimiter\ceil{\lceil}{\rceil}

\DeclareMathOperator*{\argmax}{argmax}

\def\blue{\textcolor{blue}}

\begin{document}

\preprint{}
\title{Stabilizing multiple topological fermions on a quantum computer}

\author{Jin Ming Koh}
\email{jkoh@caltech.edu}
\affiliation{Division of Physics, Mathematics and Astronomy, California Institute of Technology, Pasadena, California 91125, United States}

\author{Tommy Tai}
\affiliation{Cavendish Laboratory, University of Cambridge, JJ Thomson Ave, Cambridge CB3 0HE, United Kingdom}

\author{Yong Han Phee}
\affiliation{Department of Physics, National University of Singapore, Singapore 117542}

\author{Wei En Ng}
\affiliation{Department of Physics, National University of Singapore, Singapore 117542}
\affiliation{School of Computing, National University of Singapore, Singapore 117417}

\author{Ching Hua Lee}
\email{phylch@nus.edu.sg}
\affiliation{Department of Physics, National University of Singapore, Singapore 117542}

\begin{abstract}
In classical and single-particle settings, non-trivial band topology always gives rise to robust boundary modes. For quantum many-body systems, however, multiple topological fermions are not always able to coexist, since Pauli exclusion prevents additional fermions from occupying the limited number of available topological modes. In this work, we show, through IBM quantum computers, how one can robustly stabilize more fermions than the number of topological modes through specially designed 2-fermion interactions. Our demonstration hinges on the realization of BDI- and D-class topological Hamiltonians of unprecedented complexity on transmon-based quantum hardware, and crucially relied on tensor network-aided circuit recompilation approaches beyond conventional trotterization. We also achieved the full reconstruction of multiple-fermion topological band structures through iterative quantum phase estimation (IQPE). All in all, our work showcases how advances in quantum algorithm implementation enables NISQ-era quantum computers to be exploited for topological stabilization beyond the context of single-particle topological invariants. 
\end{abstract}

\maketitle
\date{\today}

\section{Introduction}

Many-body quantum effects like particle statistics and Hubbard interactions underscore some of the most exciting condensed matter phenomena, such as superconductivity and fractionalization~\cite{bardeen1957theory, bardeen1957microscopic, matthias1963superconductivity, orenstein2000advances, jain1989composite, wen1991non, moore1991nonabelions, ardonne1999new, bolotin2009observation, ardonne2008degeneracy, tsukazaki2010observation, lee2018floquet}. Their interplay with single-particle properties like band topology is particularly fascinating, with rich fractionalized quasiparticles emerging out of Coulomb repulsion within dispersionless Landau levels, for instance.
But unlike in purely single-body settings, particle statistics often play a central role, determining all Fermi surface properties like electrical conductivity. In topological systems, Pauli exclusion also implies that topological robustness is only conferred upon the few electrons occupying the limited number of topological modes~\cite{qi2008topological,hasan2010colloquium}.

Compared to single-body topological phenomena realizable in classical metamaterials, novel many-body phases are traditionally much harder to engineer and physically demonstrate. This is due to challenges in accessing and manipulating intrinsically fragile quantum states, unlike the macroscopic classical degrees of freedom of photonic, acoustic and electrical circuit lattices~\cite{lu2014topological,khanikaev2017two,ozawa2019topological,ota2020active,yang2015topological,fleury2016floquet,xue2019acoustic,ma2019topological,hofmann2020reciprocal,olekhno2019topological,li2019emergence,lee2020imaging,ni2020robust,lu2019probing,stegmaier2020topological,zhang2020topolectrical}. Fortunately, there is considerable recent experimental progress in various quantum systems such as ultracold atomic lattices~\cite{Bloch2012,Bernien2017}, photonic systems~\cite{Slussarenko2019}, silicon~\cite{Yang2019} and trapped ion systems~\cite{Bruzewicz2019, gyongyosi2019survey}, which make Richard Feynman's vision~\cite{Feynman1982, Boada_2015} of utilizing quantum systems to simulate quantum Hamiltonians a reality.

Of particular versatility are universal quantum simulators, also known as quantum computers, which allow arbitrary quantum systems to be simulated, thereby in principle enabling any quantum phenomenon to be physically realized. Through the appropriate design of sequences of quantum operations, known overall as quantum algorithms, quantum computers are capable of performing a vast array of computational tasks, some at polynomial or exponential resource advantage over classical algorithms~\cite{shor1994algorithms, grover1996fast, harrow2009quantum, bernstein1997quantum}. Indeed, even in the current noisy intermediate-state quantum (NISQ) era, quantum computers have already shown great promise, with demonstrations of quantum supremacy in limited settings~\cite{arute2019quantum, zhong2021quantum}, mapping topology in parameter space~\cite{roushan2014observation}, the achievement of chemical accuracy in intermediate-scale electronic structure calculations~\cite{google2020hartree}, and in neutron scattering and exotic magnetic phenomena simulations~\cite{king2021scaling, kairys2020simulating, chiesa2019quantum, francis2020quantum}.

This work shall employ computations made with IBM quantum computers, which not only rank favourably with other cutting-edge platforms~\cite{larose2019overview, blinov2021comparison} in hardware performance, such as gate fidelities and decoherence times, but is also fully accessible via the cloud. IBM Quantum (IBM Q) currently offers access to up to 65-qubit machines based on superconducting transmon qubits, and provides an open-source software development kit called \textit{Qiskit}~\cite{qiskit2020textbook, aleksandrowicz2019qiskit}. Thus far, IBM Q has been successfully utilized in simulating spin models~\cite{zhukov2018algorithmic}, global quantum quenches~\cite{smith2019simulating}, quantum chemistry problems~\cite{mcCaskey2019quantum, gard2020efficient}, topological phenomena~\cite{choo2018measurement, smith2019crossing, azses2020identification, xiao2021robust}, machine learning~\cite{havlicek2019supervised, zoufal2019quantum} and various other applications~\cite{alvarez2018quantum, wei2018efficient, behera2019designing, zhukov2018algorithmic}.

We emphasize that quantum computation in the current NISQ-era is still plagued with significant limitations. Main bottlenecks include low gate fidelity, decoherence, limited qubit connectivity and limited number of qubits~\cite{preskill2018quantum,nielsen2000quantum}, which together impose constraints on circuit depth and structure. Amidst qubit noise and readout error, typical depths of $\order{10^1}$ entangling gate layers are presently feasible for precision results. Of utmost current priority is hence the development of error mitigation and circuit optimization approaches that maximize computational capability within hardware bounds, thereby enabling the practical use of quantum simulators in contexts where classical simulators, for example electrical circuits, are inadmissible.

In this work, we push the state-of-the-art in the quantum simulation of lattice systems by stabilizing BDI- and D-class topological boundary states on a quantum computer for the first time, and demonstrating a full bandstructure reconstruction of the extended Kitaev model, also for the first time, through iterative quantum phase estimation (IQPE) for up to 2 effective fermions. Compared to existing quantum computer realizations of other topological states~\cite{choo2018measurement,smith2019crossing,azses2020identification, xiao2021robust}, some which are performed in parameter space, ours was performed on a longer (12-qubit) chain with physical open boundaries that host topological modes. Furthermore, our parent fermionic extended Kitaev Chain (KC) Hamiltonian with multiple non-local couplings presented unprecedented demands on circuit depth and complexity, necessitating the use of tensor network-aided circuit recompilation techniques beyond traditional trotterization. Importantly, we also exploited the quantum nature of the IBM machine by engineering effective interactions that serve to stabilize more fermions than originally allowed by the number of topological zero modes (up to $2$), physically realizing a novel few-body phenomenon that has not been possible in existing classical realizations of BDI- and D-class topological phases of matter. 

\section{Results}

\subsection{Physical motivation and models}

Topological robustness is a highly sought-after property exemplified  by the extraordinarily long survival duration of boundary states in specially designed topological lattices~\cite{haldane1988model,zhang2012actinide,wu2012zoology,lee2014lattice,nash2015topological,lee2017band,ezawa2018higher}. This robustness has inspired various potential applications like topological lasers, interconnects and sensors~\cite{bandres2018topological, longhi2018non,shao2020high,zhang2012chiral,liu2014spin,budich2020non}, and originates from the integer quantization of topological invariants characterizing the lattice band structure~\cite{niu1985quantized,hasan2010colloquium}. While non-trivial topology has been demonstrated in a wide variety of classical metamaterials, true quantization of the response can only be observed in quantum settings\footnote{But see Refs.~\cite{li2020topology} and \cite{leykam2021probing}.}, since classical excitations are analog signals which can be arbitrary large. 

Yet, a fully quantum topological fermionic system is also limited by the fact that there can only be as many topological fermions as available topological modes. Since the latter is determined by the topological invariant, which is typically an integer of order $\mathcal{O}(10^0)$, it will be of great scientific and practical interest to probe how quantum interactions can also \textit{enhance} the number of robustly surviving fermions beyond what is dictated by the topological invariant. We shall first introduce our topological lattice models, and later discuss how specific interactions preserve the fidelity of multiple fermions by hosting many-body states that are adiabatically connected to non-interacting topological states.

In this work we simulate, for the first time on a quantum computer, the extended Kitaev model~\cite{li2016topological, lee2020unraveling} shown in Eq.~(\ref{EKC}), which is a 1D open chain containing next-nearest neighbour (NNN) couplings in addition to its underlying nearest-neighbor (NN) structure with two sites per unit cell: 
\begin{equation}\begin{split}
    H^{\text{KC}} &= \frac{1}{2} \Bigg\{\sum_{j=1}^{N-1}\bigg[v_1 \left( c^\dagger_j c_{j+1} - d^\dagger_{j} d_{j+1} \right) + \Delta_1 \left( d^\dagger_j c_{j+1} - c^\dagger_j d_{j+1} \right)\bigg]\\
    +& \sum_{j=1}^{N-2}\bigg[v_2 \left( c^\dagger_j c_{i+2} - d^\dagger_{j} d_{j+2} \right) + \Delta_2 \left( e^{i\phi} d^\dagger_j c_{j+2} - e^{-i\phi} c^\dagger_j d_{j+2} \right) \bigg]\\
    &+\sum_{j=1}^N  \mu \left( c^\dagger_j c_j - d^\dagger_{j} d_j \right)\Bigg\}  + \text{h.c.},\label{EKC}
\end{split}\end{equation}
for $N$ unit cells, chemical potential $\mu$, tunneling coefficients $v_{1, 2}$, superconducting pairing $\Delta_{1,2}$, relative phase $\phi$, and fermionic operators $c_j, d_j$ respectively acting on the A- and B-sublattices of the $j$-th unit cell.

Our model $H^{\text{KC}}$ is an extension of the standard Kitaev chain model that has been intensely studied for hosting Majorana zero modes~\cite{kitaev2001unpaired, leumer2020exact}, which is of particular interest to fault-tolerant topological quantum computing~\cite{stern2013topological, kitaev2003fault, sarma2015majorana, lian2018topological, aasen2016milestones}. Its extra degrees of freedom in Eq.~(\ref{EKC}) allows for smooth tuning across the BDI- and D- symmetry protected topological (SPT) classes elaborated below, and will ultimately also be useful in the design of interactions that preserve the robustness of multiple fermions.

When $\phi = 0$, $H^{\text{KC}}$ preserves time-reversal symmetry (TRS) in additional to parity and charge conjugation symmetry, and belongs to the BDI class of the tenfold-way topological classification~\cite{ryu2010topological}, characterized by a winding number $\nu\in\mathbb{Z}$. Our particular model possesses $\nu = 0, 1, 2$ regimes, the first trivial and the latter two respectively exhibiting $2$-fold and $4$-fold degeneracy of midgap topological modes localized at either boundary. Mathematically, $2\pi\nu$ corresponds to the winding of its Berry phase across one Brillouin zone period (see Methods \blue{A}).

When $\phi \neq 0$, TRS is broken and $H^{\text{KC}}$ falls into the D-class characterized by $\mathbb{Z}_2$ invariant, with its topologically trivial/non-trivial phases respectively labeled by Berry phase winding $\gamma = 0, \pi$ which encapsulates the relative configurations of the $k$ and $-k$ paths~\cite{li2016topological, lee2020unraveling}. This D-class phase minimally requires NNN couplings, unlike the BDI class which contains the extremely well-known Su-Schrieffer-Heeger (SSH) model~\cite{su1979solitons, su1980soliton, bansil2016colloquium, meier2016observation, asboth2016short}, and is thus much less frequently investigated let alone realized in quantum settings~\cite{bagrets2012class}. We note that the SSH model, which we shall also simulate as an introductory example, exists as a special case of the extended Kitaev model upon unitary rotation:
\begin{equation}\begin{split}
    H^{\text{SSH}} &= \sum_{j=1}^{N} v c^\dagger_{j} d_{j} + \sum_{j=1}^{N-1}w d^\dagger_{j} c_{j+1} + \text{h.c.},\label{SSH}
\end{split}\end{equation}
with $v, w$ intra-cell and inter-cell hopping coefficients. 

\subsection{Quantum simulation of state evolution}

\begin{figure*}[!t]
    \includegraphics[width=\textwidth]{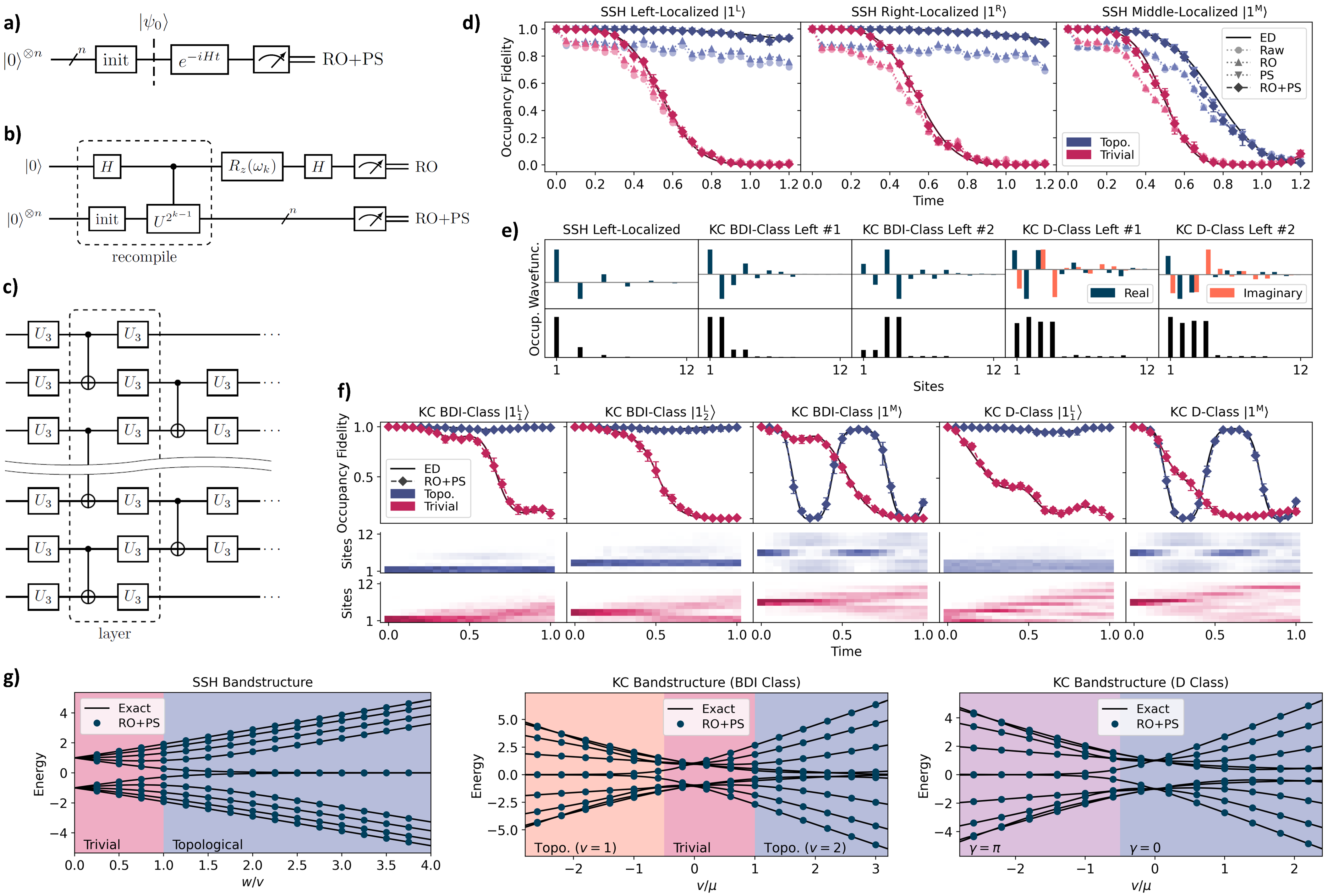}
    \caption{Quantum circuits for (a) time-evolution and (b) iterative quantum phase estimation (IQPE) for reconstructing band structure eigenergies; (c) circuit recompilation ansatz for the propagator $e^{-iHt}$, comprising an initial layer of $U_3$ gates and then successive layers of entangling CXs and single-qubit rotations. (d) $1$-particle time-evolution results for $\smash{H^{\text{SSH}}}$, with well-preserved occupancy fidelity for perfectly boundary-localized $\smash{\ket*{1^\text{L,R}}}$ states when $\nu=1$, and rapidly decaying occupancy fidelity for $\nu=0$ (non-topological) cases and all middle-localized $\smash{\ket*{1^\text{M}}}$ states. While raw hardware data already agrees qualitatively with exact diagonalization (ED) results, the agreement becomes excellent with RO and PS error mitigation. (e) Exact topological state wavefunctions at the left boundary, which constitute the dominant components of our $1$-site, $2$-site and $4$-site boundary-localized initial states detailed in Table 1. (f) $1$-particle time-evolved occupation fidelity and spatial state density for BDI-class $H^{\text{KC}}$ in topological ($\nu = 2$) and trivial ($\nu = 0$) phases; and D-class $H^{\text{KC}}$ with $\phi = \pi / 4$ in topological ($\gamma = 0$) and trivial phases. Like before, boundary localized states persist much longer when the system is topological, despite not being exact eigenstates. Middle initial states recur periodically due to boundary reflection from finite system length. (g) IQPE-obtained single-particle band structures for $\smash{H^{\text{SSH}}}$ and $\smash{H^{\text{KC}}}$, with topological phase transitions marked by changes in $2\nu$, the number of degenerate midgap ($E=0$) states. For $\smash{H^{\text{KC}}}$, parameters are held at $v_1 = v_2 = \Delta_1 = \Delta_2 \equiv v$, with $\phi = 0$ for BDI-class and $\phi = \pi/4$ for D-class (full parameter specifications in Table \ref{tab:figure-parameters}). All time-evolution results are computed on chains with $N = 6$ unit cells ($n = 12$ qubits), with error bars representing the min/max from repeated experiments across different qubit chains and machines (\textit{ibmq\_paris}, \textit{ibmq\_toronto}, \textit{ibmq\_manhattan}, and \textit{ibmq\_boeblingen}); IQPE results are for $n = 10$ chains with an additional ancilla qubit.}
    \label{fig:figure-1} 
\end{figure*}

\begin{table*}[!t]
    \renewcommand{\arraystretch}{0.45}
    \begin{tabular}{p{3.5cm} p{3.2cm} p{1.5cm} p{6cm} p{0.1cm}}
        \toprule
        Model & Regime of Stability & $\ket*{\alpha}$ & $c[\alpha]^\dagger$ & \\ 
        \midrule
        $H^{\text{SSH}}$ 
            & $\nu = 1$ & $\ket*{1^\text{L}}$ & $c^\dagger_1$  & \\ 
            & $\nu = 1$ & $\ket*{1^\text{R}}$ & $d^\dagger_{N}$  & \\
            & Non-SPT & $\ket*{1^\text{M}}$ & $c^\dagger_{\ceil{N/2}}$ & \\
        $H^{\text{KC}}$ (BDI-Class) 
            & $\nu = 1, 2$ & $\ket*{1^\text{L}_1}$ & $\frac{1}{\sqrt{2}} \big(c^\dagger_1 - d^\dagger_1\big)$  & \\ 
            & $\nu = 2$ & $\ket*{1^\text{L}_2}$ & $\frac{1}{\sqrt{2}} \big(c^\dagger_2 - d^\dagger_2\big)$  & \\ 
            & $\nu = 1, 2$ & $\ket*{1^\text{R}_1}$ & $\frac{1}{\sqrt{2}} \big(c^\dagger_{N} + d^\dagger_{N}\big)$  & \\ 
            & $\nu = 2$ & $\ket*{1^\text{R}_2}$ & $\frac{1}{\sqrt{2}} \big(c^\dagger_{N-1} + d^\dagger_{N-1}\big)$  & \\
            & Non-SPT & $\ket*{1^\text{M}}$ & $\frac{1}{\sqrt{2}} \big(c^\dagger_{\ceil{N/2}} + d^\dagger_{\ceil{N/2}}\big)$ & \\
        $H^{\text{KC}}$ (D-Class) 
            & $\gamma = 0$ & $\ket*{1^\text{L}_1}$ & $\frac{1}{2} \big(c^\dagger_1 - e^{i\phi} d^\dagger_1 + i c^\dagger_2 - i e^{i\phi} d^\dagger_2 \big)$  & \\ 
            & $\gamma = 0$ & $\ket*{1^\text{L}_2}$ & $\frac{1}{2} \big(c^\dagger_1 - e^{i\phi} d^\dagger_1 - i c^\dagger_2 + i e^{i\phi} d^\dagger_2 \big)$  & \\ 
            & $\gamma = 0$ & $\ket*{1^\text{R}_1}$ & $\frac{1}{2} \big(c^\dagger_{N-1} + e^{i\phi} d^\dagger_{N-1} + i c^\dagger_{N} + i e^{i\phi} d^\dagger_{N} \big)$  & \\ 
            & $\gamma = 0$ & $\ket*{1^\text{R}_2}$ & $\frac{1}{2} \big(c^\dagger_{N-1} + e^{i\phi} d^\dagger_{N-1} - i c^\dagger_{N} - i e^{i\phi} d^\dagger_{2N} \big)$  & \\
        \bottomrule
    \end{tabular}
    \caption{Definitions of initial $1$-particle boundary states $\smash{\ket*{\alpha} = c[\alpha]^\dagger \ket*{\text{vac}}}$, to be evolved by $H^{\text{SSH}}$ and $H^{\text{KC}}$. They include idealized edge states localized at certain sites near the boundary, as well as subscripted `M' states at the middle of the chain, used for comparison purposes. Some of these states are adiabatically connected to topological eigenstates, as elaborated in Methods \blue{A}. Multi-particle idealized states are constructed from the respective $c[\alpha]^\dagger$ operators---for example $\smash{\ket*{2^\text{LR}_{12}} = c[1^\text{L}_1]^\dagger c[1^\text{R}_2]^\dagger \ket*{\text{vac}}}$.}
    \label{tab:idealized-edge-states}
\end{table*}

We next discuss how fermionic state dynamics are studied on quantum computers. Given an initial state $\ket{\psi_0}$ and a time-independent Hamiltonian $H \in \{H^{\text{SSH}}, H^{\text{KC}}\}$, the time-evolved state is $\ket{\psi(t)} = U(t) \ket{\psi_0}$, with propagator $U(t) = e^{-iHt}$. Traditional implementation of $U(t)$ on quantum circuits entails expanding $H$ in the spin-$1/2$ basis and employing trotterization~\cite{georgescu2014quantum, ortiz2001quantum, somma2002simulating}; but acceptable trotterization error requires numerous time steps, yielding deep circuits. Furthermore, each trotter step, comprising terms $\smash{e^{-i \beta_{\vb*{\sigma}} \vb*{\sigma} \Delta t}}$ for Pauli strings $\vb*{\sigma}$ and coefficients $\beta_{\vb*{\sigma}}$ (See Methods \textcolor{blue}{C}), requires layers of entangling gates scaling with the weight of $\vb*{\sigma}$, thus impairing its usefulness for models with longer-range couplings. To transcend these limitations, we employ an implementation strategy known as circuit recompilation~\cite{sun2021quantum, khatri2019quantum, jones2020quantum, heya2018variational}. A circuit ansatz (Figure \ref{fig:figure-1}c) comprising $n_L \leq 8$ repetitive layers of $U_3$ rotation and CX entangling gates is iteratively optimized, through tensor network-based quantum simulation~\cite{gray2018quimb}, to approach the intended unitary. Specifically, the $U_3$ angles $\vb*{\chi} = (\vb*{\theta}, \vb*{\phi}, \vb*{\lambda})$ are fine-tuned with noise-robust L-BFGS-B with basin-hopping (see Methods \textcolor{blue}{C}). This technique yields order-of-magnitude shallower circuits than trotterization at comparable error rates, and is critical in our acquisition of high-quality experiment data on NISQ hardware.

A schematic of the time-evolution circuit is given in Figure \ref{fig:figure-1}b, comprising the $\ket{\psi_0}$ initialization, recompiled $U(t)$ propagator, and readout. Through computational-basis measurements on simulation qubits, the occupancy $\vb{O} = [\mqty{\expval{O_1} & \expval{O_2} & \ldots & \expval{O_n}}]^\intercal$ for $O_j = c^\dagger_j c_j = (1 - \sigma^z_j) / 2$ along the chain is retrieved. The extent of evolution away from $\ket{\psi_0}$, whose occupancy is $\vb{O}_0$, is then assessed via a metric of occupancy fidelity $0 \leq \mathcal{F}_O = \abs{\vb{O}^\intercal \vb{O}_0}^2 \leq 1$. We employ error mitigation techniques to improve the quality of hardware results---first, readout error mitigation (RO) reverses bit-flips during measurement based on prior calibration of qubits~\cite{kandala2019error, kandala2017hardware, temme2017error, qiskit2020textbook, smith2019simulating}; second, post-selection (PS) is performed on particle number~\cite{mcardle2019error, smith2019simulating}. Indeed, since $H^{\text{SSH}}$ and $H^{\text{KC}}$ are number-conserving, time-evolved states that fall outside the particle number sector of $\ket{\psi_0}$ are unphysical and can be discarded; no additional circuit depth is incurred since the measurements for $\vb{O}$ suffice. To feasibly accommodate the number of qubits $n$ used in our simulations ($7 \leq n \leq 13$), we adopt a tensored RO scheme that differs from the open-source implementation in \emph{Qiskit}~\cite{qiskit2020textbook}. See Methods \textcolor{blue}{C} for technical details of our abovementioned quantum simulation methods.

A conceptual remark pertains to the mapping between the fermionic models and the spin-$1/2$ qubits on the quantum computers. In traditional trotterization, the Jordan-Wigner or Bravyi-Kitaev transforms~\cite{georgescu2014quantum, ortiz2001quantum, somma2002simulating, nielsen2000quantum, seeley2012bravyi} map the fermionic Hamiltonian into spin chains; the constituent Pauli strings are then naturally expressible on quantum circuits. Using circuit recompilation, an approximate mapping between the fermionic and spin-$1/2$ bases still necessarily exists, but is dynamically determined during optimization.

\subsection{Persistent boundary modes from topological protection}

As the first experiments on the IBM quantum computer, we demonstrate that initial states at the boundary survive much longer when the Hamiltonian is topological. We emphasize that this is true for for a large set of boundary states, not just topological eigenstates. For a start, various perfectly localized 1-fermion initial states defined in Table \ref{tab:idealized-edge-states} are evolved via $H^{\text{SSH}}$ and $H^{\text{KC}}$.

We present $1$-particle time-evolution results for $H^{\text{SSH}}$ in Figure \ref{fig:figure-1}d, comparing raw data, data with RO and PS mitigation, and data with both. The effectiveness of the error mitigation methods is clear, with the occupation fidelity of all initial states approaching close to exact diagonalization (ED) results with RO and PS applied. The stability, \textit{i.e.} persistence, of the initial state is quantified via the occupancy fidelity $\mathcal{F}_O$; the slower the decay in $\mathcal{F}_O$, the more stable the state. As expected, the decay of the initial states $\smash{\ket*{1^\text{L}}}$, $\smash{\ket*{1^\text{R}}}$ localized at the left and right boundary sites are slow in the topological regime, compared to that of the middle-localized state $\smash{\ket*{1^\text{M}}}$, which overlaps negligibly with topological boundary modes; by contrast, in the trivial regime topological protection does not exist and all states decay quickly. 

For $H^\text{KC}$ with NNN couplings, even relatively delocalized initial states can exhibit slow decay. We determine `idealized' maximally localized boundary modes that are adiabatically connected to topological eigenstates (see Methods \blue{A}), that exhibits negligible decay, such as to facilitate the design of more general persistent states. In the BDI class, they are $\smash{\ket*{1^\text{L}_1}}$ and $\smash{\ket*{1^\text{L}_2}}$ as defined in Table \ref{tab:idealized-edge-states}, which are localized on the first and second unit cells\footnote{First pair and second pair of sites.} respectively. From Figure \ref{fig:figure-1}f, $\ket*{1^\text{L}_1}$ (and $\ket*{1^\text{R}_1}$) are robust for both $\nu=1$ and $2$ topological sectors, but $\ket*{1^\text{L}_2}$ (and $\ket*{1^\text{R}_2}$) are robust only for the $\nu=2$ sector, which supports two topological boundary modes at each end. For D-class $H^\text{KC}$ systems with nonzero $\phi$ and NNN couplings, the four idealized boundary states are each localized on the four boundary sites with $\phi$-dependent relative phases between site orbitals, as defined in Table \ref{tab:idealized-edge-states}. Indeed, almost negligible decay was observed (Figure \ref{fig:figure-1}f) for these states in the topological regimes, compared to the middle-localized $\smash{\ket*{1^{\text{M}}}}$ which do not benefit from topological robustness; in the trivial regime, all states decay quickly as expected. The extent of delocalization of $\smash{\ket*{1^\text{M}}}$ in all regimes, and of all initial states in the topologically trivial regime, are detailed in the occupancy density maps of Figure \ref{fig:figure-1}f. 

We emphasize that this marks the first time D-class topological physics, and Hamiltonians of the complexity of $H^\text{KC}$, have been accessed on digital quantum computers. While state initialization in the time-evolution circuits for $H^{\text{SSH}}$ and BDI-class $\smash{H^{\text{KC}}}$ are performed through explicit circuit components (see Methods \blue{C}), that for $\smash{\ket*{1^\text{L,R}_{1,2}}}$ of D-class $\smash{H^{\text{KC}}}$ are absorbed into our dynamically optimized recompilation ansatz, so as to minimize incurred costs in circuit depth. Testimony to the fidelity of our experiments, run on quantum hardware, is the minimal decay experienced by our topological boundary states (Figure \ref{fig:figure-1}d and especially \ref{fig:figure-1}f), which significantly exceeds that without RO and PS error mitigation techniques, with or without topological protection. 

\subsection{Full band structure reconstruction}

Our boundary states owe their slow decay to the gapped nature of their dominant (topological) eigenstate component, as is readily understood through a two-component approximate treatment. Consider $\ket{\psi(t)} = \alpha \ket*{\psi_1} + \beta \ket*{\psi_2} e^{-i \Delta E t}$ with $\abs{\alpha}^2 + \abs{\beta}^2 = 1$; then the state fidelity is $\mathcal{F}(t) = \abs{\braket{\psi(t)}{\psi(0)}}^2 = 1 - 4 \abs{\alpha}^2 \abs{\beta}^2 \sin^2{(\Delta E t / 2)}$, which stays near unity as long as $\abs{\alpha}^2 \abs{\beta}^2 \ll 1$, that is, when a dominant eigenstate component exists. This can be checked by comparing all our initial states with exact topological eigenstates (Figure \ref{fig:figure-1}e). In realistic settings where the initial boundary state generically overlaps with arbitrarily many eigenstates, the energy gap between the dominant topological eigenstate and all other states is crucial to stability. It introduces a separation of frequency scales that avoids overwhelming destructive interference. Experimentally, this energy gap can be verified from a full reconstruction of the topological band structure. Below, we first describe our approach for mapping the band structure, and then present its results for both single-fermion and two-fermion systems.

\begin{figure*}[!t]
    \includegraphics[width=\textwidth]{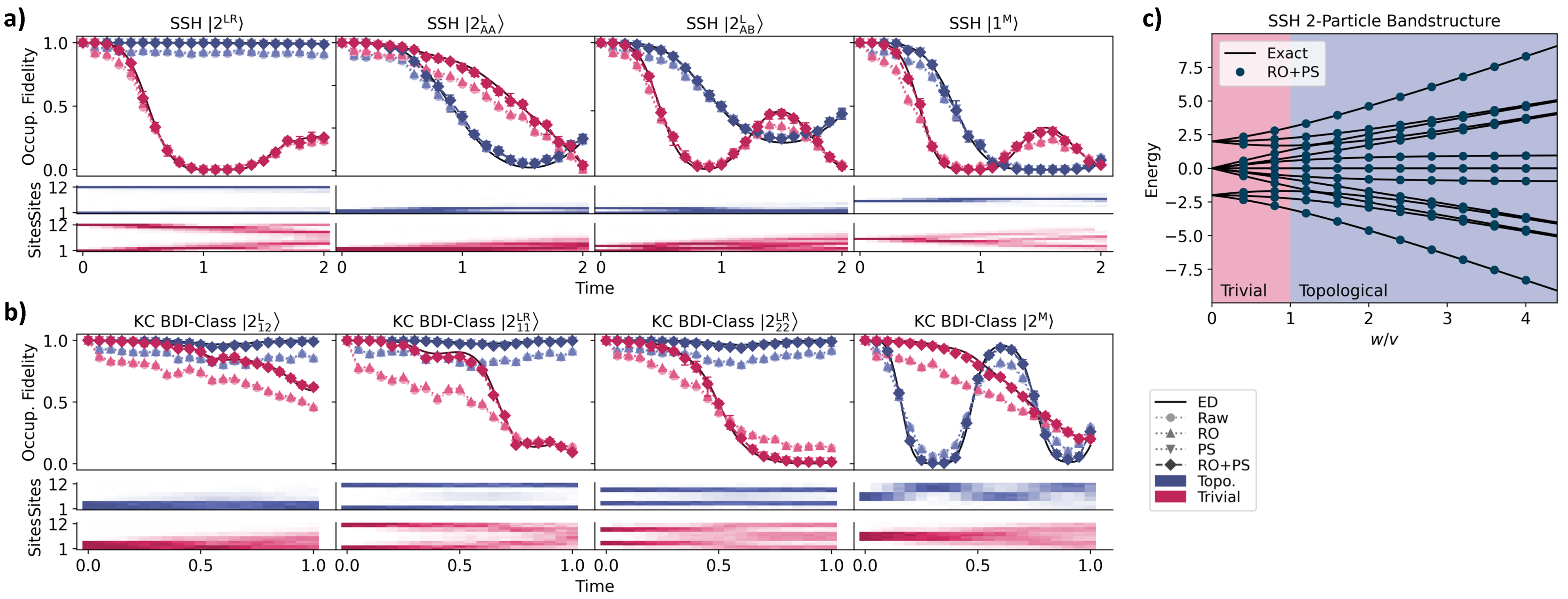}
    \caption{(a) $2$-fermion time-evolution results for $H^{\text{SSH}}$ when $\nu=1$, with occupancy fidelity preserved only for the initial state $\smash{\ket*{2^\text{LR}}}$, which is perfectly localized on both edges. The left-localized $\smash{\ket*{2^\text{L}_\text{AB}}}$ and $\smash{\ket*{2^\text{L}_\text{AA}}}$ states decay as rapidly as $\smash{\ket*{1^\text{M}}}$ regardless of topology; indeed it is impossible to construct $2$-fermion robust states localized on a single edge. (b) In contrast, $H^{\text{KC}}$ in the $\nu = 2$ phase stabilizes the left-localized $2$-fermion $\smash{\ket*{2^\text{L}}}$ state, as well as various other states localized at both edges. (c) 2-fermion band structure of $H^{\text{SSH}}$ as accessed through IQPE. All time-evolution computations are for $N = 6$ unit cell ($n = 12$ qubit) chains, with detailed parameter specifications given by Table \ref{tab:figure-parameters}; IQPE results are on $n = 7$ qubits, one of which is an ancilla. IBM Q machines \textit{ibmq\_paris}, \textit{ibmq\_toronto}, and \textit{ibmq\_manhattan} were used.}
    \label{fig:figure-2}
\end{figure*}

To probe and reconstruct the band structure on quantum hardware, we perform iterative quantum phase estimation (IQPE)~\cite{miroslav2007arbitrary, mohammadbagherpoor2019improved}. Given $U \ket{\psi} = e^{2 \pi i \phi} \ket{\psi}$ for unitary $U$ and eigenstate $\ket{\psi}$, IQPE estimates the eigenphase $\phi \in [0, 1)$, in principle to arbitrary precision. Setting $U = e^{-iHt}$ allows the inference of eigenenergy $E = - 2 \pi \phi / t$ of $\ket{\psi}$. Compared to quantum phase estimation (QPE)~\cite{whitfield2011simulation, aspuru2005simulated, cleve1998quantum, nielsen2000quantum}, IQPE circuits are shallower and requires fewer qubits---only a single ancilla qubit and a single controlled-unitary block is required. There is no need for multi-qubit inverse Fourier transforms. Truncating the binary expansion $\phi = 0.\phi_1 \phi_2 \ldots \phi_m$ to $m$ bits, IQPE iterates from $k = m$ to $k = 1$; in iteration $k$, a controlled-$\smash{U^{2^{k-1}}}$ block and a feedback $R_z(\omega_{k}) = -2\pi(0.0\phi_{k+1} \phi_{k+2} \ldots \phi_{m})$ rotation are applied, and the ancilla qubit is measured to determine $\phi_k$. An IQPE circuit diagram is shown in Figure \ref{fig:figure-1}b. To minimize circuit depth, the initialization of $\ket{\psi}$ and the controlled-unitary block are implemented via recompilation; RO mitigation is applied to all qubits, and PS is applied to the simulation qubits to select for specific particle number sectors. Indeed, by performing IQPE over numerous $\ket{\psi}$---which need not be exact eigenstates, since superposition states collapse at measurement and yield expected eigenenergies nonetheless---the band structure of the Hamiltonian $H$ can be recovered at arbitrarily high resolution. See Methods \textcolor{blue}{C} for technical details.

IQPE results for $H^{\text{SSH}}$ and $H^{\text{KC}}$ in the $1$-particle sector are shown in Figure \ref{fig:figure-1}(g). The BDI topological phase transition in $H^{\text{SSH}}$ at $w=v$ is apparent, with a pair of midgap states at separating from the bulk and approaching degeneracy at $E=0$ as $w / v$ increases. The $H^{\text{KC}}$ model possesses richer behaviour---in the BDI case, transitions from the $\nu = 1$ phase into the trivial ($\nu = 0$) and then into $\nu = 2$ phase occur as $v / \mu$ increases, for illustrative $v \equiv v_1 = v_2 = \Delta_1 = \Delta_2$. The $\nu = 1$ phase exhibits two-fold degeneracy of midgap states like in the SSH model, and the $\nu = 2$ phase exhibits four-fold degeneracy at $E=0$ due to having two zero modes at each boundary. In the D class, two-fold midgap topological degeneracy occurs in the $\gamma = \pi$ but not $\gamma = 0$ phase. For the latter, four-fold degeneracy is broken as the eigenenergies split into pairs of positive energy and negative energy states, which can be regarded as remnants of topological midgap states. In all cases, the reconstructed band structure from hardware execution of IQPE closely match exact diagonalization results.

\subsection{Topological stability for multiple fermions --- non-interacting case}

Compared to existing classical realizations of BDI- and D-class topological states, our IBM Q realizations possess the distinct advantage of accessing quantum many-body effects like fermionic statistics and interactions. We shall first discuss the former, specifically on the various ways whereby two fermions can simultaneously enjoy topological robustness. 
We first consider $H^{\text{SSH}}$, for which two-fermion boundary states can be constructed either as $\smash{\ket*{2^\text{LR}} = c[1^\text{L}]^\dagger c[1^\text{R}]^\dagger \ket*{\text{vac}}}$ with 1 fermion at each boundary, or alternatively as $\smash{\ket*{2^\text{L}_\text{AA}}}$ or $\smash{\ket*{2^\text{L}_\text{AB}}}$ with one fermion in the $\smash{\ket*{1^\text{L}}}$ orbital and the other in the nearest A or B site. Of these, only $\ket{2^\text{LR}}$ has both particles overlapping significantly with topological boundary modes, so only it would be conferred stability in the topological phase (see Figure \ref{fig:figure-2}a). The $\smash{\ket*{2^\text{L}_\text{AA}}}$ and $\smash{\ket*{2^\text{L}_\text{AB}}}$ states, like $\smash{\ket*{1^\text{M}}}$, are unstable whether topological modes exist or not, despite one of the fermions being in the topologically stable orbital $\smash{\ket*{1^\text{L}}}$. The upshot is that since $H^{\text{SSH}}$ possesses only one topological state at each end of the chain, it is impossible to construct a $2$-particle topological state localized at only one boundary. Generalizations to more fermions should be intuitive. For completeness, we also show the $2$-particle band structure from IQPE (Figure \ref{fig:figure-2}c); the topological phase is characterized by pairs of bands merging into degeneracy as $w/v \to \infty$, even though the zero energy midgap states can now correspond to either $\ket{2^\text{LR}}$ or the product state of two particles with equal and opposite eigenenergies.

\begin{table*}[!t]
    \renewcommand{\arraystretch}{0.35}
    \begin{tabular}{p{3cm} p{4cm} p{7cm} p{0.1cm}}
        \toprule
        Figure & Regime & Parameters \\ 
        \midrule
        \ref{fig:figure-1}d, \ref{fig:figure-2}a, \ref{fig:figure-3}c, \ref{fig:figure-3}d
            & SSH $\nu = 1$ & $v = 1/2$, $w = 1$. & \\
            & SSH $\nu = 0$ & $v = 3/2$, $w = 1$. & \\
        \ref{fig:figure-1}f, \ref{fig:figure-2}b
            & KC BDI-Class $\nu = 2$ & $\mu = 1$, $v_1 = \Delta_1 = 1/2$, $v_2 = \Delta_2 = 5$, $\phi = 0$. & \\
            & KC BDI-Class $\nu = 0$ & $\mu = 7$, $v_1 = \Delta_1 = 5/2$, $v_2 = \Delta_2 = 0$, $\phi = 0$. & \\
            & KC D-Class $\gamma = 0$ & $\mu = 1$, $v_1 = \Delta_1 = 1/2$, $v_2 = \Delta_2 = 5$, $\phi = \pi / 4$. & \\
            & KC D-Class $\gamma = 0^*$ & $\mu = 5$, $v_1 = \Delta_1 = 4$, $v_2 = \Delta_2 = 0$, $\phi = \pi / 4$. & \\
        \ref{fig:figure-3}e
            & KC $\nu = 2$ & $\mu = 1$, $v_1 = \Delta_1 = 1/2$, $v_2 = \Delta_2 = 5$, $\phi = 0$. & \\
            & KC $\nu = 1$ & $\mu = 1$, $v_1 = \Delta_1 = 2$, $v_2 = \Delta_2 = 1$, $\phi = 0$. & \\
            & KC $\nu = 0$ & $\mu = v_1 = \Delta_1 = 5/2$, $v_2 = \Delta_2 = 0$, $\phi = 0$. & \\
        \ref{fig:figure-3}f
            & SSH $\nu = 1$ & $v = 1/2$, $w = 1$. & \\
        \bottomrule
    \end{tabular}
    \caption{Summary of parameter values used in the various figures. $^*$This regime contains eigenstates that are adiabatically connected to the $\nu = 0$ phase as we approach BDI class ($\phi \to 0$).}
    \label{tab:figure-parameters}
\end{table*}

To realize $2$-fermion topological states localized at a \textit{single} boundary, the four-fold midgap topological modes of $\smash{H^\text{KC}}$, with two at each boundary, can be utilized. In particular, the two fermions can occupy $\smash{\ket*{1^\text{L}_{1}}}$ and $\smash{\ket*{1^\text{L}_{2}}}$ near the left boundary, resulting in $\smash{\ket*{2^\text{LL}_{12}}}$, or analogously $\smash{\ket*{2^\text{RR}_{12}}}$ for the right boundary. Relaxing the requirement of localization on a single edge, robust states such as $\smash{\ket*{2^\text{LR}_{ij}}=c[1^\text{L}_i]^\dagger c[1^\text{R}_j]^\dagger \ket*{\text{vac}}}$ are all possible, where $i,j \in {1,2}$. Indeed, hardware results indicate these states are conferred stability in the topological regime; the contrast against the non topologically protected middle-localized $\smash{\ket*{2^\text{M}}}$ is drastic (Figure \ref{fig:figure-2}b). The occupancy density maps concur that these $2$-fermion edge localized states remains persist almost perfectly in the topological phase, but diffuses in the trivial phase.

\begin{figure*}[!t]
    \includegraphics[width=\textwidth, height=14cm]{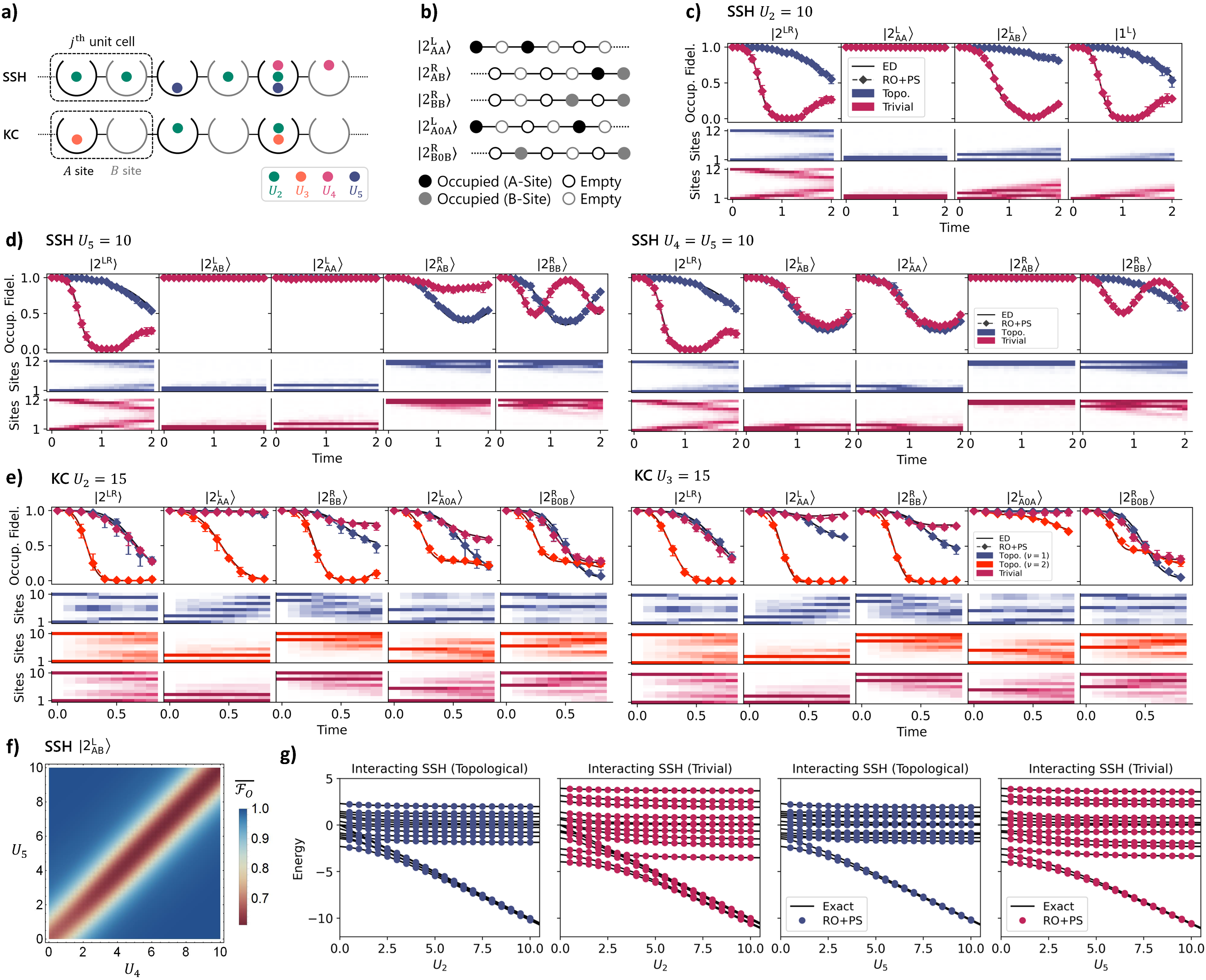}
    \caption{Schematics and $2$-particle results for the interacting models. (a) Schematic illustration of the $2$-particle Hubbard interaction terms added on top of $H^{\text{SSH}}$ and $H^{\text{KC}}$. (b) The various $2$-fermion initial boundary states that we dynamically evolve. They are named according to the sublattice occupancy of the two fermions---superscripts indicates the edges (left/right) the particles reside in, while subscripts indicate the occupied sublattices. (c-e) Time-evolution results for the $2$-fermion initial states for $H^{\text{SSH}}$ (c, d) and $H^{\text{KC}}$ (e). For $H^{\text{SSH}}$, both left (AA) and right (BB) boundary states are stabilized when $U_2 \neq 0$ only, while either AB states are preferentially stabilized when $U_5\neq 0$ only and $U_4 = U_5$. In addition, the left (AA) boundary state is stable when $U_5 \neq 0$. For $H^{\text{KC}}$, the stability of $2$-fermion boundary states competes with the topological localization when interactions $U_2$ or $U_3$ are nonzero. $U_2$ and $U_3$ preferentially stabilizes the left successive sublattice A (AA0) and next successive sublattice A (A0A) modes respectively. (f) In the $U_4$-$U_5$ parameter space, having $U_4 = U_5$ minimizes average occupancy fidelity $\overline{\mathcal{F}_O}$ of the $2$-fermion left boundary state $\smash{\ket*{2^\text{L}_\text{AB}}}$, leading to fastest decay. (g) $2$-fermion band structure of the interacting SSH model, accessed through IQPE. An additional bunch of interaction-induced bands appears below, with bandgap increasing linearly with $U_2$ and $U_5$. Time-evolution results for the interacting SSH and KC models are for chains $N = 6$ unit cells ($n = 12$ qubits) and $N = 5$ unit cells ($n = 10$ qubits) respectively, and error bars are min/max of repeated experiments; IQPE results are on $n = 7$ qubits, one of which is an ancilla. See Table \ref{tab:figure-parameters} for detailed parameter specifications.}
    \label{fig:figure-3} 
\end{figure*}

\subsection{Stability for multiple fermions --- interacting case}

Quantum interactions can present new phase transitions~\cite{PhysRevB.83.075103, PhysRevB.81.134509, Wang629, PhysRevB.82.075106,PhysRevLett.114.185701} and avenues of stability with no non-interacting analogues~\cite{Rachel_2018, Hohenadler_2013,PhysRevB.92.075135, PhysRevLett.114.185701, PhysRevLett.109.096403}. The understanding of strongly correlated topological models is also crucial in understanding the phenomenology of real materials~\cite{PhysRevLett.102.256403, PhysRevLett.109.066401, PhysRevB.83.205101, doi:10.1146/annurev-conmatphys-031214-014749, Pesin2010}. While weak interactions may only perturb the stability of SPT boundary modes, strong interactions can drastically break the symmetry protection altogether, leading to new preferred states~\cite{PhysRevB.90.165136}. Our time-evolution and IQPE methods, hinging on circuit recompilation, readily supports the study of strongly correlated models, facilitating digital quantum computers as alternative experimental platforms from existing ultracold atomic lattices~\cite{PhysRevX.7.031057, Atala2013} or photonic crystals~\cite{PhysRevLett.110.076403, PhysRevLett.109.106402}. 

We consider two-body Hubbard interactions which have been commonly used to model strong density-density interactions~\cite{PhysRevX.7.031057,PhysRevLett.110.260405,PhysRevB.26.4278, PhysRevB.97.201115, PhysRevB.86.205119, PhysRevB.82.075106}. Among the simplest 1D topological model with interactions is the SSH-Hubbard model, which has been explored in various contexts~\cite{PhysRevB.26.4278, PhysRevB.99.064105, PhysRevB.95.115443, Sirker_2014, PhysRevB.91.115118}. We are interested in Hubbard interactions that directly compete with the boundary-localization of SSH topological modes. Specifically, we consider interactions between all successive sites ($U_2$), sites within the same unit cell ($U_4$), and same sublattice sites across adjacent unit cells ($U_5$). Our interacting SSH model is then obtained by adding these interaction terms to Eq. (\ref{SSH}), $H= H^\text{SSH}+H^{\text{SSH}}_{\text{int}}$, with 
\begin{equation}\begin{split}
    H^{\text{SSH}}_{\text{int}} =& -(U_2+U_4)\sum_{j=1}^N c_j^\dag c_j d^\dag_j d_j \\
    & \, -U_2\sum_{j=1}^{N-1} c_{j+1}^\dag c_{j+1} d_j^\dag d_j
    -U_5 \sum_{j=1}^{N-1} c_{j+1}^\dag c_{j+1} c_j^\dag c_j,
    \label{intSSH}
\end{split}\end{equation}
where $c^\dagger_j$ ($d^\dagger_j$) creates a fermion in the A (B) sublattice of the $j$-th unit cell. We emphasize that these interactions are homogeneous across all unit cells, and do not induce boundary effects on their own, without the interplay with topological boundary localization. The schematic in Figure \ref{fig:figure-3}a illustrates the interaction terms in Eq. (\ref{intSSH}). The $U_4$ and $U_5$ terms are intentionally chosen to induce asymmetry across the two sublattices---they impact states with inhomogeneous polarizations differently, such as SSH topological modes which occupy one sublattice exclusively. We dynamically evolve on IBM Q various boundary-localized 2-fermion states $\smash{\ket*{2^\text{LR}}}$, $\smash{\ket*{2^\text{L}_\text{AB}}}$, $\smash{\ket*{2^\text{L}_\text{AA}}}$, $\smash{\ket*{2^\text{R}_\text{AB}}}$, $\smash{\ket*{2^\text{L}_\text{BB}}}$, schematically shown in Figure \ref{fig:figure-3}b. As before, the superscripts indicates the edge-localizations (left/right) of the 2 fermions, while the subscripts indicates the occupied sublattices; we remind the reader that the chain starts with an A-site on the left.

Our chosen $2$-particle interactions have enabled us to achieve stable multi-fermion boundary states due to either topological localization or interaction-induced polarization, and often the interplay of both. As such, we can for instance produce stable states of $H^\text{SSH}$ localized on a single boundary, reminiscent of topological states, but yet existing even in the topologically trivial regime. The conferred stability generally increases with interaction strength at diminishing returns; beyond $U \sim 10$, little additional stability is gained from stronger interactions. Yet, the impact of the topological character of the non-interacting terms usually remains significant even in the face of arbitrarily strong interactions. 

In Figures \ref{fig:figure-3}c/d, we present the evolutions of various boundary-localized 2-fermion initial states subject to strong interactions $U_2,U_4$ or $U_5$ of strength $10$. While such strong interactions may seem clearly dominant compared to the non-interacting $H^\text{SSH}$ part, the topological character of $H^\text{SSH}$ still affects the evolution of most of these states significantly. When $U_2$ is switched on (Figure \ref{fig:figure-3}c), $\smash{\ket*{2^\text{L}_\text{AA}}}$ (and equivalently $\smash{\ket*{2^\text{R}_\text{BB}}}$, not shown) retain near-perfect fidelity, even when the chain is topologically trivial. However, for the $\smash{\ket*{2^\text{LR}}}$ initial state localized at both boundaries, topology allows a much longer survival time. To obtain stable $2$-fermion states localized at a single boundary, we have to switch on $U_4$ and $U_5$; when $U_4 > 0$, both types of left boundary modes ($\smash{\ket*{2^\text{L}_\text{AB}}}$ and $\smash{\ket*{2^\text{L}_\text{AA}}}$) are stabilized. On the other hand, to obtain stable right boundary states, we require $U_5 > 0$ as well, such as to balance with $U_4$. This is reflected from the stability plot in parameter space (Figure \ref{fig:figure-3}f)---the time-averaged occupancy fidelity $\overline{\mathcal{F}_O}$ for the left-localized $2$-particle boundary states is minimal along the $U_4=U_5$ diagonal, being greatly destabilized relative to the right-localized modes. The preferential stabilization at either boundary by $U_4$ and $U_5$ is a consequence of their coupling of specific sublattice pairs on finite chains; when the interaction couples all sublattices like for $U_2$, two-fermion boundary modes are stable on both edges, and the preferential stability is not observed.

Similarly, one can study Hubbard interactions on the extended Kitaev model. We again focus on interaction terms that induce sublattice asymmetry, and thus left/right boundary asymmetry for topological modes. The extended Kitaev chain (Eq. \ref{EKC}), however, has longer-range couplings compared to the SSH model (Eq. \ref{SSH}). The NNN hoppings and pairings were crucial in the non-interacting model to demonstrate a richer topology, exhibiting more than one pair of topological modes in the $\nu = 2$ phase of the BDI class. We thus consider interactions that compete with this topology, specifically between occupied A-sites of NN ($U_2$) an NNN ($U_3$) unit cells, with $H=H^\text{KC}+H^{\text{KC}}_{\text{int}}$,
\begin{equation}\begin{split}
    H^{\text{KC}}_{\text{int}} = -U_2\sum_{j=1}^{N-1} c_j^\dag c_j c_{j+1}^\dag c_{j+1}
        - U_3 \sum_{j=1}^{N-2} c_j^\dag c_j c_{j+2}^\dag c_{j+2},
    \label{intEKC}
\end{split}\end{equation}
$H^\text{KC}$ set to be in the BDI class. These interactions are schematically illustrated in Figure \ref{fig:figure-3}a. We dynamically evolve the various 2-fermion states used earlier, $\smash{\ket*{2^\text{L}_\text{AA}}}$, $\smash{\ket*{2^\text{L}_\text{BB}}}$, $\smash{\ket*{2^\text{L}_\text{AB}}}$ and $\smash{\ket*{2^\text{R}_\text{AB}}}$, which now have large overlaps with the topological modes of the non-interacting Kitaev chain (see Figure \ref{fig:figure-3}b for schematics). Recall that the BDI-class Kitaev chain can host up to two pairs of topological modes at the two boundaries. As shown in Figure \ref{fig:figure-3}e, the presence of either interaction ($U_2$ or $U_3$) worsens the fidelity of all $2$-particle boundary modes when $\nu = 2$. These interactions hence disrupt the symmetry protection conferred in the BDI class. When $\nu = 1$, there is one fewer pair of topological boundary modes, but the states $\smash{\ket*{2^\text{L}_\text{A0A}}}$ and $\smash{\ket*{2^\text{R}_\text{B0B}}}$---which do not overlap significantly with any non-interacting topological mode---now enjoy enhanced robustness in the presence of interactions. Also, all edge-localized modes can be made robust in the topologically trivial case $\nu = 0$. These observations demonstrate how NNN Hubbard interactions can lead to new avenues of stability, which is not wholly surprising, given that they are known to lead to new phases like topological Mott phases in other contexts like honeycomb lattices~\cite{PhysRevLett.100.156401}. 

Finally, we present the full two-fermion band structure of the interacting SSH model (Eq. \ref{intSSH}) reconstructed on quantum hardware using IQPE (Figure \ref{fig:figure-3}g). With strong interactions, an additional band of energies whose large gap scales linearly with interaction strength appears. However, the interpretation of this additional set of bands is not straightforward. The interacting Hamiltonian does not share the same eigenstates as its non-interacting counterpart; in fact differences in eigenstate wavefunctions are drastic even for modest $U \sim v, w$ comparable to the non-interacting $H^\text{SSH}$. One therefore cannot hope to understand the separation of the band purely using the non-interacting eigenstates. Moreover, exact diagonalization reveals that the band does not necessarily contain $2$-fermion boundary-localized modes, even though initial $2$-fermion boundary-localized states can be stabilized in some cases. The near-perfect fidelity of these stabilized $2$-particle boundary modes, some representatives having been previously demonstrated (Figure \ref{fig:figure-3}), is due to strong overlap with one of the many eigenstates $\ket{\psi}$ of the interacting Hamiltonian; but the location of $\ket{\psi}$ in the band structure may be in either set of bands.

\section{Conclusion}
\label{sec:conclusion}

By realizing the interacting extended Kitaev Chain on IBM quantum computers, we have demonstrated how various boundary states can be robustly preserved by topological mechanisms of BDI- and D-class symmetries. Importantly, this topological protection is not limited to topological eigenstates, and interplays non-trivially with 2-body interactions and Pauli exclusion statistics in multi-fermion settings. Most spectacularly, we discovered avenues where interactions allows more fermions to be stabilized at one boundary than suggested by the number of available topological modes alone. Our work also illustrates how tensor network-aided circuit recompilation techniques beyond traditional trotterization enables the simulation and full band structure reconstruction of topological Hamiltonians of unprecedented complexity on quantum circuits. Our approach can be further extended to more sophisticated many-body interacting Hamiltonians, presenting a myriad of new opportunities and raising the state-of-the-art in the quantum simulation of strongly correlated topological systems.

\section{Acknowledgements}

J. M. Koh thanks Shi-Ning Sun of Caltech for helpful discussions on quantum computing and algorithms. We acknowledge the use of IBM Quantum services for this work. The views expressed are those of the authors, and do not reflect the official policy or position of IBM or the IBM Quantum team.

\clearpage
\onecolumngrid

\appendix

\section{Physical Models}
\label{app:models}

\subsection{Single-particle topology: Su-Schrieffer-Heeger (SSH) model and extended Kitaev chains (KCs)}
\label{app:models/ssh-kc}

Here we review the symmetry-protected topological (SPT) phases realized in this work. In general, SPT phases are gapped phases protected by at least one global symmetry---for example time reversal, charge conjugation or crystalline symmetry~\cite{senthil2015symmetry, chiu2016classification, haldane2017nobel, wen2017colloquim}. In the recent decades, the study of SPT phases has been greatly fueled by the discovery of topological insulators, topological superconductors and topological semimetals~\cite{ryu2010topological,hasan2010colloquium,asboth2016short, bernevig2013topological}. Each topological phase is characterized by at least one topologically invariant quantity. A topological phase transition occurs when this invariant changes upon varying any of the parameters of the system. The simplest of such invariants takes the form of winding numbers in momentum space, as we shall describe below.

In this work, we consider the BDI- and D-class SPT phases in 1D, which are characterized by topological invariants that are defined at the level of single particle states. Our phases are realized via the extended Kitaev chain (KC) model, which encompasses the BDI class with $\mathbb{Z}$ invariant, as epitomized by the well-known Su-Schrieffer-Heeger (SSH) model, and a less well-known D class with $\mathbb{Z}_2$ topological invariant. 

The archetypal Su-Schrieffer–Heeger (SSH) model takes the form of a spinless Hermitian chain with two sublattice sites A and B per unit cell~\cite{su1979solitons, su1980soliton, bansil2016colloquium, meier2016observation, asboth2016short}. This model has been extensively realized in various platforms, particularly classical ones~\cite{meier2016observation,St-Jean2017a, Lee2018,wang2019topologically,downing2019topological}. The inter-site hopping depends on whether the sites are in the same sublattice or not:
\begin{equation}\begin{split}
    H^{\text{SSH}} &= \sum_{j=1}^{N} v c^\dagger_{j} d_{j} + \sum_{j=1}^{N-1}w d^\dagger_{j} c_{j+1} + \text{h.c.},
    \label{app-SSH}
\end{split}\end{equation}
for a chain with $N$ unit cells under open boundary conditions (OBCs), with intracell and intercell hopping coefficients $v$ and $w$ respectively. When $\abs{w} > \abs{v}$, this chain supports topologically robust edge states with zero eigenenergies, as can be intuitively visualized in a mechanics setting~\cite{lee2018topological}. To understand this topological robustness mathematically, we expand $H^{\text{SSH}}$ in the Pauli basis, such that it takes the form of a 3-vector $\vb{h}(k)$ in momentum space:
\begin{equation}
   H^{\text{SSH}}(k)= \vb{h}(k) \cdot \vb*{\sigma} = h_x(k) \sigma^x + h_y(k) \sigma^y + h_z(k) \sigma^z,
\end{equation}
where $\vb{h}(k)$ traces a closed loop due to the periodicity of lattice momentum $k$. Importantly, due to the sublattice symmetry characteristic of the BDI class, $h_z(k)=0$ and $\vb{h}(k)$ is restricted to a 2D plane. Its topology is hence classified by the first homotopy group $\mathbb{Z}$ of a punctured plane, \textit{i.e.} the number of times $\vb{h}(k)$ encircles the origin, which is nonzero when the chain is topologically non-trivial.

To motivate the parent model we used in this work, namely the extended Kitaev chain (KC), we note that the SSH model is mathematically equivalent to the simple Kitaev chain (SKC) developed for explaining Majorana zero modes in topological superconductivity~\cite{kitaev2001unpaired}. In its original form, its Hamiltonian contains hopping terms ($t$) and Bogoliubov-de Gennes pairing ($\Delta$) terms between adjacent sites, as well as on-site chemical potential ($\mu$) terms:
\begin{equation}
    H^{\text{SKC}} = \sum_j \left[ t \left( \eta^\dagger_j \eta_{j+1} + \eta^\dagger_{j+1} \eta_j \right) + \Delta \left( \eta^\dagger_j \eta^\dagger_{j+1} + \eta_{j+1} \eta_j \right) \right] + \mu \sum_j \eta^\dagger_j \eta_j,
    \label{eq:ham-SKC-common}
\end{equation}
where $\eta^\dagger,\eta$ creates/annihilates the Bogoliubov-de Gennes quasiparticles. It can be shown via a Bogoliubov transformation to the $c,d$ operators that $H^{\text{SKC}}$ is unitarily equivalent to the SSH chain~\cite{verresen2017one}, with the chemical potential term corresponding to an effective coupling between the two SSH sublattices~\cite{leumer2020exact}. Importantly, SSH topological boundary modes correspond to highly sought-after Majorana zero modes at the ends of these chains, whose potential application in fault-tolerant quantum computing has led to intense investigations~\cite{jiang2011majorana,lobos2012interplay,aasen2016milestones,van2020photon}, and Ref.~\cite{bagrets2012class} for D class.

By introducing next nearest neighbour (NNN) hoppings and pairings~\cite{li2015winding,li2014topological}, this model can be generalized into an \emph{extended} Kitaev chain model with versatile topological properties. In the basis containing sublattices A and B, the Hamiltonian takes the form
\begin{equation}\begin{split}
    H^{\text{KC}} &= \mu \sum_{j=1}^N \left( c^\dagger_j c_j - d^\dagger_{i} d_j \right) 
    + \frac{v_1}{2} \sum_{j=1}^{N-1} \left( c^\dagger_j c_{j+1} - d^\dagger_{i} d_{j+1} + \text{h.c.} \right)
    + \frac{v_2}{2} \sum_{j=1}^{N-2} \left( c^\dagger_j c_{j+2} - d^\dagger_{i} d_{j+2} + \text{h.c.} \right) \\
    &\qquad + \frac{i \Delta_2 \sin{\phi}}{2} \sum_{j=1}^{N-2} \left( c^\dagger_j d_{j+2} + d^\dagger_j c_{j+2} - \text{h.c.} \right) + \frac{\Delta_2 \cos{\phi}}{2} \sum_{j=1}^{N-2} \left( d^\dagger_j c_{j+2} - c^\dagger_j d_{j+2} + \text{h.c.} \right) \\
    &\qquad + \frac{\Delta_1}{2} \sum_{j=1}^{N-1} \left( d^\dagger_j c_{j+1} - c^\dagger_j d_{j+1} + \text{h.c.} \right),
    \label{ham-EKC}
\end{split}\end{equation}
where operators $c_j, d_j$ respectively act on the A- and B-sites of the $j$-th unit cell. Removing NNN hoppings, phases and pairings ($v_2 = \phi=\Delta_2 = 0$) recovers the standard 2-band Kitaev chain (SKC), or equivalently the SSH model. In momentum space, Eq.~(\ref{ham-EKC}) takes the form~\cite{li2015winding} $H^{\text{KC}}(k) = \vb{h}^\text{KC} \cdot \vb*{\sigma}$: 

\begin{align}
    h_x^\text{KC}(k) &= \Delta_2 \sin{\phi} \sin{2k}, \nonumber\\
    h_y^\text{KC}(k) &= \Delta_2 \cos{\phi} \sin{2k} + \Delta_1 \sin{k}, \nonumber\\
    h_z^\text{KC}(k) &= \mu - v_1 \cos{k} - v_2 \cos{2k}.
    \label{eq:ham-EKC-momentum}
\end{align}
The competition between nearest neighbour and next nearest neighbour terms results in a richer phase diagram, supporting up to two topological zero modes at each end. These topological modes are evident in the bandstructure. Notably, when $\phi$ is not a multiple of $\pi$, all three components of $\vb{h}^\text{KC}$ are nonzero, and the KC model possesses D-class symmetry characterized by a $\mathbb{Z}_2$ topological invariant, according to the ten-fold-way classification~\cite{ryu2010topological, schnyder2008classification}. But when $\phi$ is a multiple of $\pi$, $ h_x^\text{KC}(k)$ vanishes, and  the model only belongs to the BDI class characterized by a $\mathbb{Z}$ topological invariant, the same as the SSH model. In the winding number perspective, the topological invariant of the BDI class is the number of times $\vb{h}(k)$ encircle the origin. Again, when $\phi$ is a multiple of $\pi$, $\vb{h}(k)$ is constrained to a plane. However, when $\phi$ is not a multiple of $\pi$, $\vb{h}(k)$ traces out a closed path on a unit sphere. While there is no longer a well-defined winding around the origin, the path still possesses chiral symmetry which forces its $k$ and $-k$ paths to be related. The geometric interpretations of its two $\mathbb{Z}_2$ classes are elaborated in Refs.~\cite{lee2020unraveling} and \cite{li2019geometric}, albeit in a more general non-Hermitian context.

We note that all Hamiltonians considered in our work conserves total particle number as a symmetry, that is,

\begin{equation}\begin{split}
    \comm{H^{\text{KC}}}{O} = \comm{H^{\text{SSH}}}{O} = 0, \qquad
     O = \sum_{j = 1}^N \left(c_j^\dagger c_j + d_j^\dagger d_j\right),
\end{split}\end{equation}
hence enabling our use of post-selection on particle number (see Appendix \ref{app:error/ps}).

\subsection{Idealized Edge Modes}
\label{app:models/edge-modes}

Under certain extreme parameter limits, the topological edge mode wavefunctions of $H^{\text{SSH}}$ and $H^{\text{KC}}$ take particularly simple forms. We refer to these as idealized edge states. The robustness conferred by their SPT nature means that away from these limits, topological edge modes still possess strong overlap with these idealized ones, so long as the system remains in the same topological phase. We summarize these idealized states in Table \ref{tab:app-idealized-edge-states}. They are most conveniently given in the form $\ket{\alpha} = c[\alpha]^\dagger \ket{\text{vac}}$ for creation operator $\smash{c[\alpha]^\dagger}$ and vacuum state $\ket{\text{vac}} = \ket{000\ldots}$.

\begin{table}[!t]
    \renewcommand{\arraystretch}{0.45}
    \begin{tabular}{p{3.5cm} p{3.2cm} p{1.5cm} p{7cm} p{0.1cm}}
        \toprule
        Model & Regime of Stability & $\ket*{\alpha}$ & $c[\alpha]^\dagger$ & \\ 
        \midrule
        $H^{\text{SSH}}$ 
            & $\nu = 1$ & $\ket*{1^\text{L}}$ & $c^\dagger_1$  & \\ 
            & $\nu = 1$ & $\ket*{1^\text{R}}$ & $d^\dagger_{N}$  & \\ 
        $H^{\text{KC}}$ (BDI-Class) 
            & $\nu = 1, 2$ & $\ket*{1^\text{L}_1}$ & $\left(c^\dagger_1 - d^\dagger_1\right) \Big/ \sqrt{2}$  & \\ 
            & $\nu = 2$ & $\ket*{1^\text{L}_2}$ & $\left(c^\dagger_2 - d^\dagger_2\right) \Big/ \sqrt{2}$  & \\ 
            & $\nu = 1, 2$ & $\ket*{1^\text{R}_1}$ & $\left(c^\dagger_{N} + d^\dagger_{N}\right) \Big/ \sqrt{2}$  & \\ 
            & $\nu = 2$ & $\ket*{1^\text{R}_2}$ & $\left(c^\dagger_{N-1} + d^\dagger_{N-1}\right) \Big/ \sqrt{2}$  & \\ 
        $H^{\text{KC}}$ (D-Class) 
            & $\gamma = 0$ & $\ket*{1^\text{L}_1}$ & $\left(c^\dagger_1 - e^{i\phi} d^\dagger_1 + i c^\dagger_2 - i e^{i\phi} d^\dagger_2 \right) \Big/ 2$  & \\ 
            & $\gamma = 0$ & $\ket*{1^\text{L}_2}$ & $\left(c^\dagger_1 - e^{i\phi} d^\dagger_1 - i c^\dagger_2 + i e^{i\phi} d^\dagger_2 \right) \Big/ 2$  & \\ 
            & $\gamma = 0$ & $\ket*{1^\text{R}_1}$ & $\left(c^\dagger_{N-1} + e^{i\phi} d^\dagger_{N-1} + i c^\dagger_{N} + i e^{i\phi} d^\dagger_{N} \right) \Big/ 2$  & \\ 
            & $\gamma = 0$ & $\ket*{1^\text{R}_2}$ & $\left(c^\dagger_{N-1} + e^{i\phi} d^\dagger_{N-1} - i c^\dagger_{N} - i e^{i\phi} d^\dagger_{2N} \right) \Big/ 2$  & \\
        \bottomrule
    \end{tabular}
    \caption{Idealized topological edge states of $H^{\text{SSH}}$ and $H^{\text{KC}}$ corresponding to extreme parameter limits in Table \ref{tab:parameter-limits}.}
    \label{tab:app-idealized-edge-states}
\end{table}

\begin{table}[!t]
    \begin{tabular}{p{3.5cm} p{1.5cm} p{7cm}}
        \toprule
        Model & Regime & Parameter Limit \\ 
        \midrule
        $H^{\text{SSH}}$ 
            & $\nu = 1$ & $v / w \to 0$ \\
        $H^{\text{KC}}$ (BDI-Class) 
            & $\nu = 1$ & $\mu, t_2, \Delta_2, \phi \to 0$, $t_1 = \Delta_1$ \\
            & $\nu = 2$ & $\mu, t_1, \Delta_1, \phi \to 0$, $t_2 = \Delta_2$ \\
        $H^{\text{KC}}$ (D-Class) 
            & $\gamma = \pi$ & $\mu, t_2, \Delta_2 \to 0$, $t_1 = \Delta_1$ \\
            & $\gamma = 0$ & $\mu, t_1, \Delta_1 \to 0$, $t_2 = \Delta_2$ \\
        \bottomrule
    \end{tabular}
    \caption{Summary of extreme parameter limits of $H^{\text{SSH}}$ and $H^{\text{KC}}$.}
    \label{tab:parameter-limits}
\end{table}

The idealized edge states $\smash{\ket*{1^\text{L,R}_1}}$ of the $\gamma = \pi$ phase of D-class $H^{\text{KC}}$ are identical to those of the BDI-class $H^{\text{KC}}$, which makes them redundant to investigate. From the 1-particle $\ket{\alpha}$ of our models, $2$-particle idealized states can be constructed by sequentially applying the operators of \textit{different} $1$-particle state---due to Pauli exclusion, we cannot have $2$ particles in the same state. For example, $\smash{\ket*{2^\text{LR}_{12}} = c[1^\text{L}_1]^\dagger c[1^\text{R}_2]^\dagger \ket{\text{vac}}}$. The extreme parameter limits for our models are summarized in Table \ref{tab:parameter-limits}. One can verify the $\ket{\alpha}$ listed above by checking that the associated number operators are symmetries of the corresponding Hamiltonians $\smash{H \in \{H^{\text{SSH}}, H^{\text{KC}}\}}$ in the parameter limits,
\begin{equation}\begin{split}
    \comm{H}{c[\alpha]^\dagger c[\alpha]} = 0,
\end{split}\end{equation}
which means their occupancy profiles are conserved throughout time-evolution.

\section{Quantum Computing Preliminaries}
\label{app:prelim}

We provide a brief introduction to quantum computing and simulation, and to the IBM Q platform~\cite{garcia2020ibm, cross2018ibm, qiskit2020textbook, aleksandrowicz2019qiskit} through which we run experiments on quantum hardware. The IBM Q platform is accessible through the cloud---in fact a number of machines are freely available for public use---and has been utilized in a considerable number of studies since launch~\cite{smith2019simulating, kandala2019error, kandala2017hardware, temme2017error, alvarez2018quantum, behera2019designing, mcCaskey2019quantum, zhukov2018algorithmic, sun2021quantum, motta2020determining}. To set the technical context for our methods (see Appendix \ref{app:methods}), we discuss certain specifics of quantum computing in relation to the IBM Q platform; we point readers to other articles for a more general overview~\cite{georgescu2014quantum, ortiz2001quantum, somma2002simulating, nielsen2000quantum}.

\subsection{IBM Q Platform}
\label{app:prelim/ibmq}

We describe the workflow of running experiments on IBM Q, whose details, in particular the construction of quantum circuits and the readout of measurement results, are relevant to our computation methods; though the concepts are broadly general. One uses the \textit{Qiskit} API~\cite{qiskit2020textbook, aleksandrowicz2019qiskit} on Python to define quantum circuits and submit them to IBM Q backends (machines) for execution; the backend then reports raw measurement results, which can be processed to recover expectation values, correlation functions, and other quantities of interest. The backends execute the submitted experiments asynchronously in a fairshare queuing system. We use a group of QV-$32$ backends, namely \textit{ibmq\_paris} (27 qubits), \textit{ibmq\_toronto} (27 qubits), and \textit{ibmq\_manhattan} (65 qubits), and the QV-$16$ backend \textit{ibmq\_boeblingen} (20 qubits). The quantum volume (QV) can be treated as a rough indicator of the overall capability of the machine---number of qubits, gate error rates, and decoherence times~\cite{cross2019validating, moll2018quantum}. These are the highest-volume machines accessible to the authors at the time of writing. To provide a ballpark measure of performance, the relaxation time $T_1$, that is the timescale of thermal decay of a qubit from the $\ket{1}$ state to the $\ket{0}$ state, and the dephasing time $T_2$, that is the timescale over which phase information of a qubit is lost, ranges $\SI{60}{\micro\second} \leq T_1 \approx T_2 \leq \SI{120}{\micro\second}$ on average for these machines; and typical gate time is $T_\text{g} \approx \SI{440}{\nano\second}$. Typical single-qubit ($\sqrt{X}$) gate error is $\order{10^{-3}}$, and typical two-qubit ($\text{CX}$) gate error is $\order{10^{-2}}$. See Figure \ref{fig:trott-recomp-comp}b for an illustration of qubit layout and Section \ref{app:error/chain} for the estimation and selection of low-error qubit chains.

\subsection{Quantum Gates \& Circuits}
\label{app:prelim/gates-circuits}

A qubit is a two-level quantum system; current IBM Q hardware use transmon qubits~\cite{garcia2020ibm, cross2018ibm, qiskit2020textbook} on chips cooled to cryogenic temperatures. Essentially, the qubit is an LC oscillator with a Josephson junction as the inductive element; the anharmonicity from the nonlinear inductance allows control over the ground and first excited state of the oscillator through microwave pulses. Just as classical logic gates operate on bits, quantum gates operate on qubits. First, we define the computational basis,
\begin{equation}\begin{split}
    \ket{0} = \mqty[1 \\ 0], \qquad
    \ket{1} = \mqty[0 \\ 1],
\end{split}\end{equation}
and the multi-qubit state notation $\ket{x_1 x_2 \ldots x_n}$ is shorthand for $\ket{x_1} \otimes \ket{x_2} \otimes \ldots \otimes \ket{x_n}$, where $\otimes$ is the tensor product. Note that the qubits are always distinguishable,  unlike the logical fermions that they can represent. The single-qubit Pauli gates are
\begin{equation}\begin{split}
    \sigma^x = X = \mqty[0 & 1 \\ 1 & 0], \qquad
    \sigma^y = Y = \mqty[0 & -i \\ i & 0], \qquad
    \sigma^z = Z = \mqty[1 & 0 \\ 0 & -1],
\end{split}\end{equation}
which flip a qubit around the $x$-, $y$-, and $z$- axes of the Bloch sphere. For instance, the $X$ gate is understood as the quantum analogue of the NOT classical gate, since $X\ket{0}=\ket{1}$ and $X\ket{1}=\ket{0}$. Generated from the Pauli gates are the rotation gates
\begin{equation}\begin{split}
    R^x(\theta) &= e^{-i\theta \sigma^x / 2} = \mqty[
        \cos(\theta/2)    & -i \sin(\theta/2) \\
        -i \sin(\theta/2) & \cos(\theta/2)], \\
    R^y(\theta) &= e^{-i\theta \sigma^y / 2} = \mqty[ 
        \cos(\theta/2)    & -\sin(\theta/2) \\
        \sin(\theta/2) & \cos(\theta/2)], \\
    R^z(\theta) &= e^{-i\theta \sigma^z / 2} = \mqty[ 
        e^{-i\theta/2} & 0 \\
        0              & e^{i\theta/2}].
\end{split}\end{equation}

Other commonly used gates are the Hadamard, root-$X$ and $S$ gates,
\begin{equation}\begin{split}
    H = \frac{1}{\sqrt{2}} \mqty[1 & 1 \\ 1 & -1], \qquad
    \sqrt{X} = \mqty[1 + i & 1 - i \\ 1 - i & 1 + i], \qquad
    S = \sqrt{Z} = \mqty[1 & 0 \\ 0 & i].
\end{split}\end{equation}

The $H$ gate is useful as it creates superpositions, in that $H\ket{0} = \ket{+} = \smash{\left(\ket{0} + \ket{1}\right)/\sqrt{2}}$ and $H\ket{1} = \ket{-} = \smash{\left(\ket{0} - \ket{1}\right)/\sqrt{2}}$, which are orthogonal on the equator of the Bloch sphere. That is, it shifts between the computational basis ($z$-basis) and the $x$-basis $\ket{\pm}$, which is also the single-qubit Fourier basis. All of these gates can be represented as special cases of the $U_3$ general single-qubit gate,
\begin{equation}
    U_3(\theta, \phi, \lambda) = \mqty[
            \cos(\theta/2)           & -e^{i\lambda} \sin(\theta/2) \\
            e^{i\phi} \sin(\theta/2) & e^{i(\phi + \lambda)} \cos(\theta/2)].
\end{equation}

Analogous to the $U_3$ gate, there are more restrictive $U_1$ and $U_2$ gates,
\begin{equation}
    U_2(\phi, \lambda) = U_3(\pi / 2, \phi, \lambda) = \frac{1}{\sqrt{2}} \mqty[
            1         & -e^{i\lambda} \\
            e^{i\phi} & e^{i(\phi + \lambda)}], \qquad
    U_1(\lambda) = U_3(0, 0, \lambda) = \mqty[
            1         & 0 \\
            0 & e^{i \lambda}],
\end{equation}
which may allow lower error and faster execution time on quantum hardware. Indeed, on some IBM Q machines~\cite{qiskit2020textbook}, $U_1$ gates are performed with software frame changes and therefore take negligible gate time; $U_2$ gates require frame changes and a single microwave pulse; and $U_3$ gates require two pulses. Since all operations are unitary, those that are also Hermitian are their own inverses; they are involutory. For example, $X^2 = Y^2 = Z^2 = H^2 = I$. 

We now discuss multi-qubit gates. Of relevance is the controlled-$U$ gate, which applies unitary $U$ to a target qubit if the control qubit is $\ket{1}$. For two qubits, 
\begin{equation}\begin{split}
    \text{CU}_{12} = \mqty[I & 0 \\ 0 & U] = \mqty[
            1 & 0 & 0 & 0 \\
            0 & 1 & 0 & 0 \\
            0 & 0 & U_{11} & U_{12} \\
            0 & 0 & U_{21} & U_{22}], \qquad
    \text{CU}_{21} = \mqty[
            1 & 0 & 0 & 0 \\
            0 & 0 & U_{11} & U_{12} \\
            0 & 0 & 1 & 0 \\
            0 & 0 & U_{21} & U_{22}],
\end{split}\end{equation}
where the first and second indices refer to the control and target qubits respectively. Setting $U = X$ gives the controlled-NOT (CNOT) gates, $\text{CX}_{12}$ and $\text{CX}_{21}$, which are the entangling gates on IBM Q hardware. In general, for $n$ qubits, the controlled unitary can be written
\begin{equation}\begin{split}
    \text{CU}_{ij} = \ket{0}_i \bra{0}_i \otimes I_j + \ket{1}_i \bra{1}_i \otimes U_j
    = \bigotimes_{k=1}^n \left(\ket{0}\bra{0}\right)^{\delta_{ik}} + 
        \bigotimes_{k=1}^n \left(\ket{1}\bra{1}\right)^{\delta_{ik}} U^{\delta_{jk}}, 
\end{split}\end{equation}
where padding with identity matrices on all other qubits $k \neq i, j$ is implicit for the left expression. We likewise extend Pauli gates to multi-qubit systems by padding over the remaining qubits,
\begin{equation}\begin{split}
    \sigma^\mu_j = I_{2^{j-1}} \otimes \sigma^\mu \otimes I_{2^{n-j}}, \qquad
    \mu \in \{x, y, z\}.
\end{split}\end{equation}

A quantum circuit~\cite{nielsen2000quantum, georgescu2014quantum} represents a series of gate operations on qubits. Analogous to classical Boolean circuits, wires in quantum circuits represent qubits, and time flows from the left to the right. The gates specified on the circuit are applied to the qubits in order---note, for sake of clarity, that this implies a reversal of the ordering of gates when written. For instance, $X Y Z \ket{\psi_0}$ means gates $Z$ acts on $\ket{\psi_0}$, followed by $Y$ then $X$. See Figure \ref{fig:circuits} for examples of circuit diagrams. It is standard convention that qubits are initialized to $\ket{0}$ at the start of a circuit. For organizational purposes, qubits may be grouped into registers. In \textit{Qiskit}, quantum circuits comprise quantum registers each containing a number of qubits, and classical registers to which measurement results on qubits are written~\cite{qiskit2020textbook, aleksandrowicz2019qiskit}. Quantum measurements are discussed in Section \ref{app:prelim/measurements}. 

In principle, one is free to add arbitrary gates to a circuit---arbitrary unitaries spanning numerous qubits. However, digital quantum computers are engineered to support only a handful of basis gates. For example, IBM Q machines support $\{U_1, U_2, U_3, \text{CX}\}$; in January 2021 some machines have switched to the $\{R^z, \sqrt{X}, X, \text{CX}\}$ basis set~\cite{qiskit2020textbook}. These basis sets are universal, which allows arbitrary unitaries to be approximated to any desired precision. However, while the optimal decompositions for general one- and two-qubit unitaries are known, the latter through the KAK decomposition~\cite{kraus2001optimal,vatan2004optimal}, the optimal decomposition for $n$-qubit unitaries is not fully known and non-trivial to work out~\cite{drury2008constructive, mottonen2004quantum, vartiainen2004efficient, khaneja2001cartan}. Finding circuit implementations of desired unitaries is a key challenge in programming digital quantum computers---it is for this purpose that trotterization and circuit recompilation techniques are used (see Section \ref{app:circuit}). As an ending remark, we note $U_3(\theta, \phi, \lambda) = R^z(\phi + \pi) \sqrt{X} R^z(\theta + \pi) \sqrt{X} R^z(\lambda)$ up to global phase, hence our circuit ansatz for recompilation (Section \ref{app:circuit/recompilation}), represented using $U_3$ gates, can be equivalently applied for both of the IBM Q basis sets.

\subsection{Measurements}
\label{app:prelim/measurements}

All measurements are performed in the computational basis on the backends, that is, the measured observable is $\sigma^z_i$ on qubit $i \in \mathbb{N}$. A value of $1$ is written to the designated classical bit if a qubit state of $\ket{1}$ is projectively measured; and a value of $0$ is written if $\ket{0}$ is measured~\cite{qiskit2020textbook, aleksandrowicz2019qiskit}. Since the results of quantum measurements are probabilistic, to obtain statistical estimates of expectation values many repetitions have to be performed---IBM Q enables each circuit to be run for a number of shots $n_\text{s}$ to facilitate this. At the termination of the circuit, the bit values in the classical registers are concatenated into a measurement \textit{bitstring}; the backend ultimately reports measurement results $\{s, c_s\}$ for measurement bitstrings $s \in \{0, 1\}^m$ and cumulative counts $c_s$ for $m$ classical bits. The probability of measuring $s$ is accordingly
\begin{equation}\begin{split}
    p(s) = \frac{c_s}{\sum_{s' \in \{0, 1\}^m} c_{s'}} \equiv \frac{c_s}{n_\text{s}}.
\end{split}\end{equation}

Given the measurement counts, the expectation value $\expval{\sigma^z_i}$ can be calculated as
\begin{equation}\begin{split}
    \expval{\sigma^z_i} &= \mel{\psi}{\sigma^z_i}{\psi} = p_i(0) - p_i(1), \qquad
    p_i(b) = \frac{1}{n_\text{s}} \sum_{\substack{s \in \{0, 1\}^m \\ s_i = b}} c_s,
\end{split}\end{equation}
where $\ket{\psi}$ is the quantum state before measurement and $p_i(b)$ is the probability of qubit $i$ yielding a measurement bit $b$, and we take the measurement of qubit $i$ to be written to classical bit $j$ for generality. The notation $s_j$ denotes digit $j$ of the bitstring $s$. Through basis transformations implemented in the circuit, these computational-basis measurements can be used to recover different quantities of interest.

\section{Quantum Computing Methods}
\label{app:methods}

\subsection{Quantum Algorithms}
\label{app:methods/algo}

In this section, we describe our implementation of the time-evolution of states and iterative quantum phase estimation on digital quantum computers. All experiments on quantum hardware are run through the IBM Q platform.

\subsubsection{Time Evolution}
\label{app:methods/algo/evolve}

\begin{figure*}[!t]
    \includegraphics[width=\textwidth]{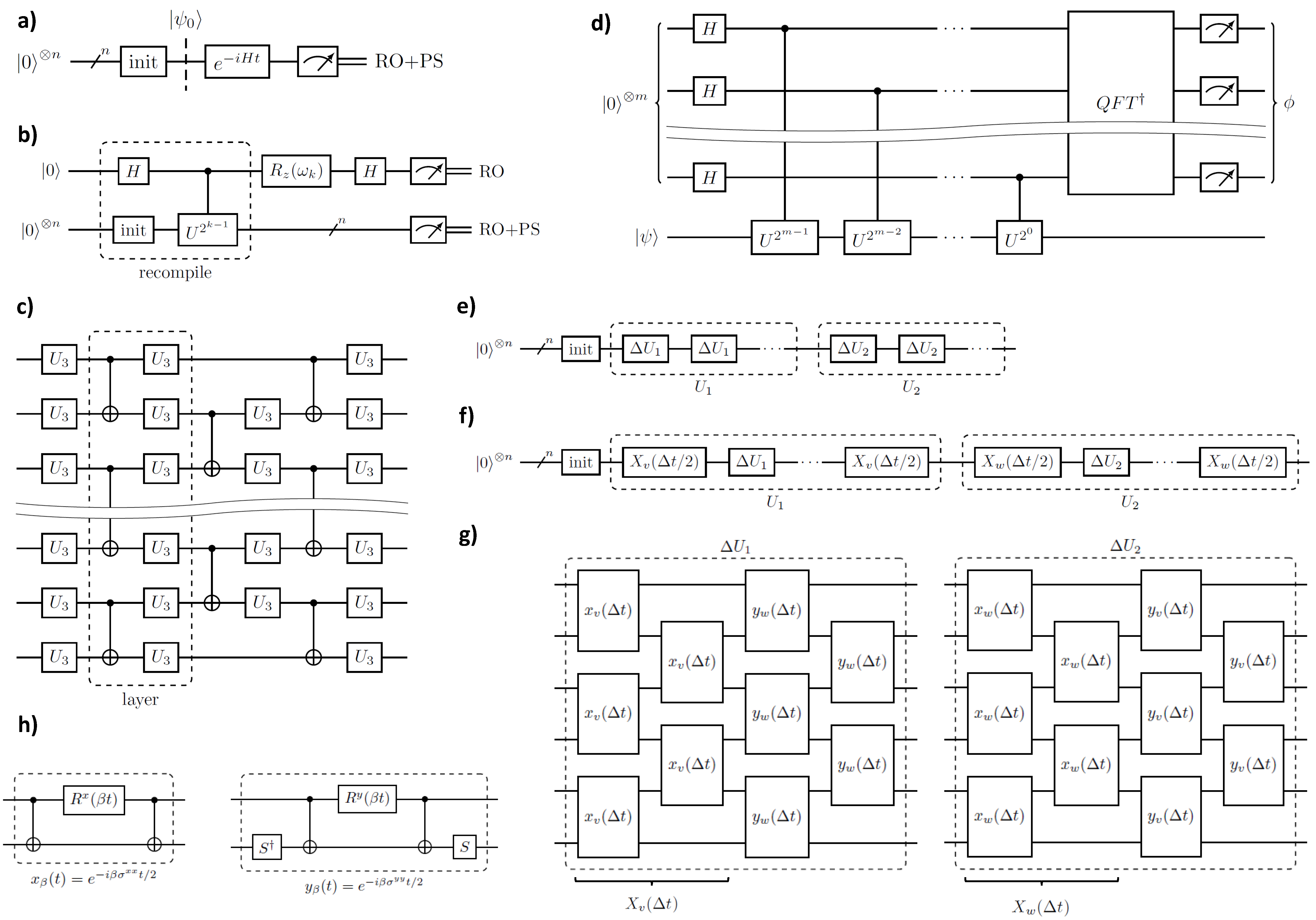}
    \caption{Summary of quantum circuits and circuit components. (a) Time-evolution and (b) iterative quantum phase estimation (IQPE) circuit schematics. (c) Circuit ansatz for recompilation. (d) Circuit for traditional quantum phase estimation (QPE). The $m$ ancilla qubits accommodates the reconstruction of the binary representation of the eigenphase $\phi$. Circuit schematics for (a) first-order trotterization and (b) second-order symmetric trotterization of $H^{\text{SSH}}$, showing the concatenated $U_1$ and $U_2$ components (see Section \ref{app:circuit/trotterization} for breakdown) and their constituent trotter steps; (g) structure of each trotter step, comprising even-odd parallelized $e^{-i \beta \sigma^{xx} t / 2}$ and $e^{-i \beta \sigma^{yy} t / 2}$ blocks; (h) implementation of aforementioned exponentiated Pauli strings.}
    \label{fig:circuits} 
\end{figure*}

The Schrödinger equation describes the time-evolution of quantum states under a Hamiltonian $H$,
\begin{equation}\begin{split}
    \dv{\ket{\psi(t)}}{t} = - i H \ket{\psi(t)},
\end{split}\end{equation}
where we have set $\hbar = 1$. Given an initial quantum state $\ket{\psi_0}$, the time-evolved state at time $t$ is given by $\ket{\psi(t)} = e^{-iHt} \ket{\psi_0}$ for time-independent $H$. Note the propagator $e^{-iHt}$ is unitary since $H$ is Hermitian in our models, and hence can be implemented on quantum circuits. This can be done through trotterization (Section \ref{app:circuit/trotterization}), which entails decomposing $H$ in the spin-$1/2$ basis and performing a truncated expansion, or through a circuit recompilation procedure (Section \ref{app:circuit/recompilation}), which entails optimizing gate parameters on an circuit ansatz to approach the propagator. In this work, we have mostly employed circuit recompilation, which we optimize to handle the complexity of our Hamiltonians. The quantum circuit implementing time-evolution then consists of the initialization of $\ket{\psi_0}$, the $e^{-iHt}$ block, and desired measurements---see Figure \ref{fig:circuits}(c) for a circuit diagram. The initialization of $\ket{\psi_0}$ is summarized in Table \ref{tab:state-initialization}.

Our interest is in accessing the occupancy expectation $\expval{O_i} = \smash{\expval*{c^\dagger_i c_i}}$ at each site of $\ket{\psi(t)}$, and the fidelity of $\ket{\psi(t)}$ relative to $\ket{\psi_0}$, to probe the time-evolution of states under $H$. The former reports on the spatial localization of states, and the latter reports on the speed of time-evolution away from $\ket{\psi_0}$. To measure $\expval{O_i}$ for sites $i \in \{1, 2, \ldots, n\}$, we perform computational basis measurements on all $n$ simulation qubits, since $\smash{c^\dagger_i c_i = (1 - \sigma^z_i) / 2}$ from the Jordan-Wigner transform (see Section \ref{app:circuit/mapping}). This applies regardless of the circuit generation technique used---trotterization or circuit recompilation---since we map single-qubit $\ket{0}$ to the unoccupied and $\ket{1}$ to the occupied fermionic states. We remind that $\sigma^z \ket{0} = \ket{0}$ and $\sigma^z \ket{1} = -\ket{1}$, therefore $\smash{\expval*{c^\dagger_i c_i} = 0}$ when qubit $i$ is projectively measured in the $\ket{0}$ state and $\smash{\expval*{c^\dagger_i c_i} = 1}$ when in $\ket{1}$, as indeed expected. Suppose the measurement counts are $c_s$ for measurement bitstring $s \in S = \{0, 1\}^n$. The recovery of $\expval{\sigma^z_i}$ from measurement results has been previously discussed (see Section \ref{app:prelim/measurements}); it can be seen that
\begin{equation}\begin{split}
    \expval{O_i} = \frac{1 - \expval{\sigma^z_i}}{2} 
    = \sum_{\substack{s \in S \\ s_i = 1}} c_s \Bigg/ \sum_{s \in S} c_s.
\end{split}\end{equation}

The set of measurements $\expval{\sigma^z_i}$ is cheap in circuit depth and enables access to $\expval{O_i}$ as desired; but does not report on the relative phase between the basis states $\ket{s}$, needed to compute the state fidelity $\smash{\mathcal{F}(\psi(t), \psi_0) = \abs{\braket{\psi_0}{\psi(t)}}^2}$. To measure $\mathcal{F}(\psi(t), \psi_0)$ quantum state tomography is required~\cite{vogel1989determination, james2005measurement, lvosky2009continuous, qiskit2020textbook}, which requires a number of circuits scaling exponentially with $n$, and basis changes for measuring the different Pauli strings that incur a considerable number of entangling gate layers. This is prohibitively expensive for large $n$. Instead of measuring $\mathcal{F}(\psi(t), \psi_0)$, we use an alternate fidelity measure $\mathcal{F}_O(\psi(t), \psi_0)$, termed the occupancy fidelity and defined generically as
\begin{equation}\begin{split}
    \mathcal{F}_O(\psi_1, \psi_2) = \abs{\vb{O}_1^\dagger \vb{O}_2}^2,
\end{split}\end{equation}
where $\vb{O}_1 = \smash{\big[\mqty{\expval{O_1}_1 & \expval{O_2}_1 & \ldots \expval{O_n}_1}\big]^\intercal}$ is the occupancy vector for $\ket{\psi_1}$ and likewise for $\vb{O}_2$. The form of the occupancy fidelity closely resembles that of the state fidelity; it serves the same purpose of reporting on the extent of evolution away from $\ket{\psi_0}$, based on occupancy expectations instead of the full quantum state. Since we readily measure $\expval{O_i}$, the occupancy fidelity $\mathcal{F}_O(\psi(t), \psi_0)$ can be accessed at no additional cost circuit-wise; that the occupancy vectors are $\order{n}$ large also means the calculation is classically efficient. Note error mitigation procedures (see Section \ref{app:error}) are applied to the measurement counts $c_s$ before recovering $\expval{\sigma^z_i}$ and $\mathcal{F}_O(\psi(t), \psi_0)$ in our experiments.

\subsubsection{Iterative Quantum Phase Estimation}
\label{app:methods/algo/iqpe}

Given $U \ket{\psi} = e^{2 \pi i \phi} \ket{\psi}$ for a unitary $U$ and an eigenstate $\ket{\psi}$, (traditional) quantum phase estimation (QPE) allows the measurement of eigenphase $\phi \in [0, 1)$, in principle to arbitrary precision~\cite{whitfield2011simulation, aspuru2005simulated, cleve1998quantum, nielsen2000quantum}. Note that the eigenvalues of a unitary have unity modulus, so there is no loss of generality. Setting $U = e^{-iHt}$ then allows the inference of eigenenergy $E = - 2 \pi \phi / t$ of $\ket{\psi}$ from the eigenphase, enabling the probing of the bandstructure of $H$. A standard circuit diagram for QPE is shown in Figure \ref{fig:circuits}(d). Notably, each ancilla qubit measures a single bit of the binary representation of $\phi$, hence to measure $\phi$ to reasonable ($<5\%$) precision requires $\order{5}$ ancillae, in addition to the $n$ simulation qubits. Furthermore, with the controlled-unitaries entangling each ancilla with the simulation qubits, and the inverse Fourier transform to shift from the Fourier basis into the computational basis for readout, means that QPE circuits are deep.

To reduce circuit breadth and depth, we use \textit{iterative} quantum phase estimation (IQPE), which has only a single ancilla qubit and controlled-unitary block~\cite{miroslav2007arbitrary, mohammadbagherpoor2019improved}; the inverse Fourier transform for a single qubit is simply a Hadamard gate. Truncating the binary expansion of $\phi = 0.\phi_1 \phi_2 \ldots \phi_m$ to $m$ bits, we iterate from $k = m$ to $k = 1$, measuring from the least to the most significant bit. In the circuit for the $k = m$ iteration, a controlled-$\smash{U^{2^{m-1}}}$ block is applied, and the ancilla qubit is measured to determine $\phi_m$. It can be shown that the probability $p_0$ of measuring an ancilla state of $\ket{0}$ is $\cos^2{\left[(0.\phi_m)\pi\right]}$ in the absence of noise, which is unity for $\phi_m = 0$ and zero for $\phi_m = 1$; of course, in the presence of noise measuring $\phi_m$ is no longer strictly deterministic, but the inference of $\phi_m = 0$ if $p_0 > 1/2$ and $\phi_m = 1$ otherwise can still be applied. Subsequently, in iteration $k$, a controlled-$\smash{U^{2^{k-1}}}$ block is applied, and a feedback $R_z(\omega_{k})$ rotation is applied to rotate off the phase due to the previous bits, before likewise inferring $\phi_{k}$. The feedback angle is $\omega_k = -2\pi(0.0\phi_{k+1} \phi_{k+2} \ldots \phi_{m})$. Clearly, to measure $\phi$ to $m$ bits of precision, $m$ iterations are needed. A circuit diagram of the IQPE circuit is given in Figure \ref{fig:circuits}(b); we perform circuit recompilation (Section \ref{app:circuit/recompilation}) to implement the initialization of $\ket{\psi}$ and the controlled-unitary block. Note readout error mitigation (Section \ref{app:error/ro}) is applied to all qubits, and post-selection is applied to the simulation qubits to select a specific number sector (Section \ref{app:error/ps}).

There are several points to note. Firstly, it is not strictly necessary that the input state $\ket{\psi}$ be an eigenstate of $U$; an arbitrary $\ket{\psi}$ can always be written in the eigenbasis of $U$ and superposition collapse during measurement will produce an eigenphase $\phi$ associated with an eigenstate of $U$. In principle, repeating QPE with randomly generated $\ket{\psi}$ will eventually record all eigenphases $\phi$ and hence the bandstructure of $H$. However, in the presence of noise, it is beneficial to choose $\ket{\psi}$ with good overlap with an eigenstate of $U$, especially in IQPE---the issue is that near-equal overlaps with two or more eigenstates could result in $p_0 \approx 1/2$ in an iteration, and an incorrect inference would produce an erroneous $\phi$. Secondly, the $2\pi$ periodicity of $\phi$ implies $2\pi\phi + 2\pi\ell = -E t$ for any $\ell \in \mathbb{Z}$, so the mapping $\phi \to E$ is multi-valued and one cannot infer a unique $E$ from a single QPE instance. To infer $E$ correctly one may choose a sufficiently small $t$ to force $\ell = 0$, and to decide the sign of $E$ one examines the slope between $\phi$ and $t$. To estimate the appropriate range of $t$, it is helpful to estimate the largest $\abs{E}$ from the form of $H$. 

To overcome these inherent limitations, we provide our IQPE instances with initial guesses $\{(\ket{\psi}, E)\}$ from exact diagonalization. This is classically tractable since we are interested in specific particle number sectors, in particular the $1$-particle and $2$-particle sectors, whose Hilbert spaces are not exponentially large with number of sites $n$. Indeed, the $1$-particle and $2$-particle subspaces are only of dimensionality $\smash{\order{n}}$ and $\smash{\order{n^2}}$; that is, the restricted Hamiltonian matrices are $\smash{\order{n \times n}}$ and $\smash{\order{n^2 \times n^2}}$ large. Since classical diagonalization is polynomially efficient in the size of the matrix, computing initial guesses $\smash{\{(\ket{\psi}, E)\}}$ takes polynomial time and space. To be clear, the $q$-particle restricted Hamiltonian matrix $H^{[q]}$ is $H$ with only rows and columns corresponding to $q$-particle basis states retained. In fact, the following construction can be found for the $1$-particle sectors,
\begin{equation}\begin{split}
    \left(H^\text{SSH}\right)^{[1]} = \mqty[
        E_0         & T_0         & \text{}   & \text{}     & \text{}     & \text{} \\
        T_0^\dagger & E_0         & T_0       & \text{}     & \text{}     & \text{} \\
        \text{}     & T_0^\dagger & E_0       & T_0         & \text{}     & \text{} \\
        \text{}     & \text{}     & \ddots    & \ddots      & \ddots      & \text{} \\
        \text{}     & \text{}     & \text{}   & T_0^\dagger & E_0         & T_0 \\
        \text{}     & \text{}     & \text{}   & \text{}     & T_0^\dagger & E_0 ], \qquad
    \left(H^\text{KC}\right)^{[1]} = -\frac{1}{2} \mqty[
        E_1          & T_1         & T_2          & \text{}     & \text{}      & \text{} \\
        T_1^\dagger  & E_1         & T_1          & T_2         & \text{}      & \text{} \\
        T_2^\dagger  & T_1^\dagger & E_1          & T_1         & T_2          & \text{} \\
        \text{}      & \ddots      & \ddots       & \ddots      & \ddots       & \ddots \\
        \text{}      & \text{}     & T_2^\dagger  & T_1^\dagger & E_1          & T_1 \\
        \text{}      & \text{}     & \text{}      & T_2^\dagger & T_1^\dagger  & E_1 ], \qquad
\end{split}\end{equation}
\begin{equation*}\begin{split}
    E_0 = \mqty[0 & v \\ v & 0], \qquad
    T_0 = \mqty[0 & 0 \\ w & 0], \qquad
    E_1 = 2 \mqty[\mu & 0 \\ 0 & \mu], \qquad
    T_1 = \mqty[v_1 & \Delta_1 \\ -\Delta_1 & -v_1], \qquad
    T_2 = \mqty[v_2 & \Delta_2 e^{i\phi} \\ -\Delta_2 e^{-i\phi} & -v_2],
\end{split}\end{equation*}
with basis in descending binary order by convention, $\{\ket{100\ldots}, \ket{010\ldots}, \ldots, \ket{\ldots 001}\}$, and similar constructions can be given for the $2$-particle sectors. Then, given eigenvector $\smash{[\mqty{\alpha_1 & \alpha_2 & \ldots & \alpha_N}]^\intercal}$ of $H^{[q]}$ with eigenenergy $E$, and supposing the corresponding basis states of $H^{[q]}$ are $\{\ket{s_1}, \ket{s_2}, \ldots, \ket{s_N}\}$ for fermionic occupancy bitstrings $s_i$, we may recover the $\ket{\psi}$ in the full Hilbert space of $H$ as
\begin{equation*}\begin{split}
    \ket{\psi} = \sum_{i = 1}^N \alpha_i \left[\prod_{j = 1}^n \big(c^\dagger_j\big)^{s_{ij}}\right] \ket{\text{vac}}.
\end{split}\end{equation*}

Furthermore, we exploit additional optimizations to enhance computational efficiency. Firstly, as noted earlier, the input states $\ket{\psi}$ need not precisely be an eigenstate of $U$; this implies the diagonalization of $H^{[q]}$ need not be performed with full precision, and the diagonalization can be computed more quickly with relaxed numerical error thresholds and less precise data types. In fact, it is not necessary to compute new initial guesses for every IQPE instance---prior guesses can be re-used, so long as they belong in a close-by neighborhood in parameter space. Also, in non-interacting cases, the $2$-particle basis states are exactly the product of $1$-particle states; then one needs only diagonalize $H^{[1]}$ to produce initial guesses for both sectors.

\subsection{Circuit Generation}
\label{app:circuit}

Time-evolution and IQPE circuits require the implementation of the $U(t) = e^{-iHt}$ propagator. While topological Hamiltonians associated with winding number invariants are most conveniently written in fermionic second-quantized forms, qubits are inherently two-state systems. To represent our topological Hamiltonians and perform time-evolution, there is hence a need to convert between the fermionic to the spin-$1/2$ Pauli basis.

\subsubsection{Jordan-Wigner transformation}
\label{app:circuit/mapping}

For pedagogical purposes, we first present the well-known Jordan-Wigner (JW) transformation that has traditionally been used for mapping between fermions and spin-$1/2$s~\cite{georgescu2014quantum, ortiz2001quantum, somma2002simulating, nielsen2000quantum}; note that in this work, most of our quantum circuits are generated via the more sophisticated circuit recompilation technique described later (Section \ref{app:circuit/recompilation}). One writes

\begin{equation}\begin{split}
    c^\dagger_j = \left( \prod_{k=1}^{j-1} \sigma^z_k \right) \sigma_j^{-}, \qquad
    c_j = \left( \prod_{k=1}^{j-1} \sigma^z_k \right) \sigma_j^{+}, \\
\end{split}\end{equation}
where $\smash{\sigma^\pm_j = \left( \sigma^x_j \pm i \sigma^y_j \right) / 2}$ are the Pauli raising and lowering operators, and number operator $n_k = c_k^\dagger c_k = (1 - \sigma_k^z) / 2$. To avoid confusion, note this Pauli basis is completely different from that used in representing the 2-component lattices in Section \ref{app:models}. The occupancy-dependent phase factors in front of $\smash{c^\dagger_j}$ and $c_j$, called strings, are necessary to enforce the fermionic anti-commutation relations. Indeed, one can check that this transformation preserves $\smash{\acomm*{c^\dagger_j}{c_k} = \delta_{jk}}$ and $\smash{\acomm*{c^\dagger_j}{c^\dagger_k} = \acomm*{c_j}{c_k} = 0}$ for all site pairs $(j, k)$. By applying the above transform, one can decompose any fermionic quadratic tight-binding Hamiltonian in the Pauli (spin-1/2) basis,
\begin{equation}\begin{split}
    H = \sum_{i, j = 1}^n t_{ij} c^\dagger_i c_j + \text{h.c.} = \sum_{\vb*{\mu}} \alpha_{\vb*{\mu}} \sigma^{\vb*{\mu}}, \qquad
    \sigma^{\vb*{\mu}}_j = \prod_{k = 1}^{\abs{\vb*{\mu}}} \sigma^{\vb*{\mu}_k}_{j + k - 1},
\end{split}\end{equation}
where $\sigma^{\vb*{\mu}}$ are products of Pauli matrices with $\vb*{\mu}_j \in \{x, y, z\}$, and $t_{ij} \in \mathbb{C}$ and $\alpha_{\vb*{\mu}} \in \mathbb{C}$ are coefficients. As explicit examples, we have
\begin{equation}\begin{split}
    H^{\text{SSH}} &= \underbrace{\frac{v}{2} \sum_{j = 1}^N \sigma^{xx}_{2j - 1}}_{X_v} 
        + \underbrace{\frac{v}{2} \sum_{j = 1}^N \sigma^{yy}_{2j - 1}}_{Y_v} 
        + \underbrace{\frac{w}{2} \sum_{j = 1}^{N - 1} \sigma^{xx}_{2j}}_{X_w} 
        + \underbrace{\frac{w}{2} \sum_{j = 1}^{N - 1} \sigma^{yy}_{2j}}_{Y_w}, \\
    H^{\text{KC}} &= \frac{\mu}{2} \sum_{j = 1}^N \left( 
            \sigma^{z}_{2j} - \sigma^{z}_{2j-1} \right)
        + \frac{v_1}{4} \sum_{j = 1}^{N - 1} \left( 
            \sigma^{xzx}_{2j-1} + \sigma^{yzy}_{2j-1} - \sigma^{xzx}_{2j} - \sigma^{yzy}_{2j} \right) \\
        & \quad + \frac{v_2}{4} \sum_{j = 1}^{N - 2} \left( 
            \sigma^{xzzzx}_{2j-1} + \sigma^{yzzzy}_{2j-1} - \sigma^{xzzzx}_{2j} - \sigma^{yzzzy}_{2j} \right) \\
        & \quad + \frac{\Delta_1}{2} \sum_{j = 1}^{N - 1} \left( 
            \sigma^{xx}_{2j} + \sigma^{yy}_{2j} - \sigma^{xzzx}_{2j - 1} - \sigma^{yzzy}_{2j} \right)
        + \frac{\Delta_2 e^{i\phi}}{4} \sum_{j = 1}^{N - 2} \left( 
            \sigma^{xx}_{2j} + \sigma^{yy}_{2j} - \sigma^{xzzx}_{2j - 1} - \sigma^{yzzy}_{2j} \right)
\end{split}\end{equation}
where we have labelled the terms in $H^{\text{SSH}}$ by $X_{v, w}$ and $Y_{v, w}$ for convenience in the following subsection.

We remark that the JW-transform is not the only choice for mapping between the fermionic and spin-$1/2$ bases. It is a characteristic that the JW-transform stores fermionic occupancy information locally, but delocalizes parity information across all previous qubits---this manifests in the string phase factors. The parity map is opposite, storing parity locally but delocalizing occupancy across all previous qubits; and the Bravyi-Kitaev (BK) transformation~\cite{seeley2012bravyi} is a middle ground, delocalizing both parity and occupancy information, but with logarithmic scaling. In principle it is less costly circuit-wise to implement fermionic operators on a quantum computer using the BK scheme, but expected benefits at $\order{10}$ qubits for our purposes are negligible, and there is incurred complexity with applying the necessary update operations. Hence the JW-transform can often be the preferred choice, by virtue of its simplicity.

\subsubsection{Trotterization}
\label{app:circuit/trotterization}

Decomposing the Hamiltonian in the Pauli basis does not lead immediately to a circuit implementation of the $U(t) = e^{-iHt}$ propagator; the problem is that the Pauli strings summed over may not commute. It is known that $e^{A + B} \neq e^A e^B$ when $\comm{A}{B} \neq 0$ for operators $A$ and $B$, hence one cannot simply implement the exponentiated Pauli strings one-by-one and glue the circuits together. This problem is addressable with the Suzuki-Trotter (ST) decomposition~\cite{vidal2004efficient, trotter1959product, sieberer2019digital, heyl2019quantum, smith2019simulating, sun2021quantum}. The main observation is that
\begin{equation}\begin{split}
    e^{A+B} = \lim_{n\to\infty} \left( e^{A/n} e^{B/n} \right)^n,
\end{split}\end{equation}
hence we break $\smash{e^{-iHt} = \left( e^{-iH \Delta t} \right)^M}$ for $\Delta t = t / M$ into $M \gg 1$ pieces, called trotter steps, each of which can be expanded to small error. For $H = \sum_{\vb*{\mu}} \alpha_{\vb*{\mu}} \sigma^{\vb*{\mu}}$, we have the first-order trotterization
\begin{equation}\begin{split}
    e^{-i H \Delta t} = \left(\prod_{\vb*{\mu}} e^{-i \alpha_{\vb*{\mu}} \sigma^{\vb*{\mu}} \Delta t}\right) + \order{\frac{1}{M^2}},
\end{split}\end{equation}
and hence the total error incurred over $M$ steps is $\order{1/M}$, suppressible by increasing $M$. Each of the $\smash{e^{-i \alpha_{\vb*{\mu}} \sigma^{\vb*{\mu}} \Delta t}}$ terms can be straightforwardly implemented on a quantum circuit, since $\sigma^{\vb*{\mu}}$ is a Pauli string. There are higher-order ST decompositions as well~\cite{hatano2005finding, smith2019simulating}. Splitting $H = \sum_{\vb*{\mu}} \alpha_{\vb*{\mu}} \sigma^{\vb*{\mu}} + \sum_{\vb*{\nu}} \alpha_{\vb*{\nu}} \sigma^{\vb*{\nu}}$, the second-order symmetric trotterization can be written as
\begin{equation}\begin{split}
    e^{-i H \Delta t} = \left(\prod_{\vb*{\mu}} e^{-i \alpha_{\vb*{\mu}} \sigma^{\vb*{\mu}} \Delta t / 2}\right)
    \left(\prod_{\vb*{\nu}} e^{-i \alpha_{\vb*{\nu}} \sigma^{\vb*{\nu}} \Delta t}\right)
    \left(\prod_{\vb*{\mu}} e^{-i \alpha_{\vb*{\mu}} \sigma^{\vb*{\mu}} \Delta t / 2}\right)
    + \order{\frac{1}{M^3}},
\end{split}\end{equation}
hence yielding total error $\order{1/M^3}$. The symmetry of the decomposition allows gate layers to be combined between adjacent trotter steps, reducing circuit depth.

We illustrate and compare the first- and second-order trotterized circuits for the SSH model in Figure \ref{fig:circuits}(e)--(h). Gates acting on different qubits are parallelized whenever possible to reduce circuit depth; this is most efficiently done by splitting the sums into odd and even terms. For the second-order symmetric trotterization, we note $\comm{X_v}{X_w} = \comm{Y_v}{Y_w} = 0$ and $\comm{X_v}{Y_v} = \comm{X_w}{Y_w} = 0$, and so $H^{\text{SSH}}$ splits into two commuting groups $X_v + Y_w$ and $X_w + Y_v$. Accordingly, we may write the propagator $U(t) = e^{-iHt} = U_2(t) U_1(t)$, for $U_1(t) = e^{-i(X_v + Y_w)t}$ and $U_2(t) = e^{-i(X_w + Y_v)t}$, yielding the trotterization structure shown. Notice that the implementation of $\smash{e^{-i \beta \sigma^{xx} t / 2}}$ and $\smash{e^{-i \beta \sigma^{yy} t / 2}}$ unitaries require 2 CX layers each; in general the necessary CX depth increases with the weight of the Pauli string.

\subsubsection{Circuit Recompilation}
\label{app:circuit/recompilation}

Traditional trotterization results in deep circuits for all but the simplest Hamiltonians---consider that the time-evolution unitaries $U_\text{SSH}(t)$ and $U_\text{KC}(t)$ require $4$ and $>16$ CX layers per trotter step. Ballpark estimates of $T_1 \approx T_2 \approx \SI{60}{\micro\second}$ and CX gate time of $\SI{440}{\nano\second}$ suggest a maximum feasible circuit depth of ${\sim}140$ CX layers; in practice hardware results degrade noticeably after ${\sim}10$ CX layers, without accounting for the presence of other gates in the circuit. It is advantageous to identify more compact circuits serving in the same capacity. This motivates the circuit recompilation procedure~\cite{sun2021quantum, khatri2019quantum, jones2020quantum, heya2018variational} discussed here, which we employ for our computations on IBM Q hardware.

We adopt a circuit ansatz comprising an initial layer of $U_3$ gates on all qubits followed by $n_L$ ansatz layers, each comprising a layer of CX gates entangling adjacent qubits and a layer of $U_3$ single-qubit rotations (see Figure \ref{fig:circuits}c). This ansatz provides sufficient freedom for the unitaries in this work at modest $n_L \leq 8$ circuit depth. Each $U_3$ gate in the ansatz is associated with angles $(\theta, \phi, \lambda)$; we collate all of them into parameter vector $\vb*{\theta}$. Then, given a target circuit unitary $V$ and an initial state $\ket{\psi_0}$, we numerically treat the optimization problem
\begin{equation}\begin{split}
    \argmax_{\vb*{\theta}} F(V_{\vb*{\theta}} \ket{\psi_0}, V \ket{\psi_0})
    = \argmax_{\vb*{\theta}} \abs{\mel{\psi_0}{V_{\vb*{\theta}}^\dagger V}{\psi_0}}^2,
\end{split}\end{equation}
where $V_{\vb*{\theta}}$ is the circuit ansatz unitary with parameters $\vb*{\theta}$. Then the recompiled circuit is the ansatz with optimal parameters $\vb*{\theta}^*$ fixing the $U_3$ gates. In our implementation, we use a tensor network-based noiseless circuit simulator~\cite{gray2018quimb} to compute $V_{\vb*{\theta}}$ and the limited-memory Broyden–Fletcher–Goldfarb–Shanno algorithm with box constraints (L-BFGS-B) to perform the optimization~\cite{andrew2007scalable, malouf2002comparison}, with basin hopping for $10^2$ hops and at most $10^3$ iterations per hop. The optimization is terminated at target fidelity $F \geq 99.99\%$. This procedure produces recompiled circuits typically in seconds to minutes; to further aid efficiency we store all recompiled circuits in a persistent cache for rapid reuse. Note the recompilation procedure works for arbitrary circuit unitary $V$---so long as $n_L$ provides sufficient freedom for fitting.

\begin{figure*}[!t]
    \includegraphics[width=\textwidth]{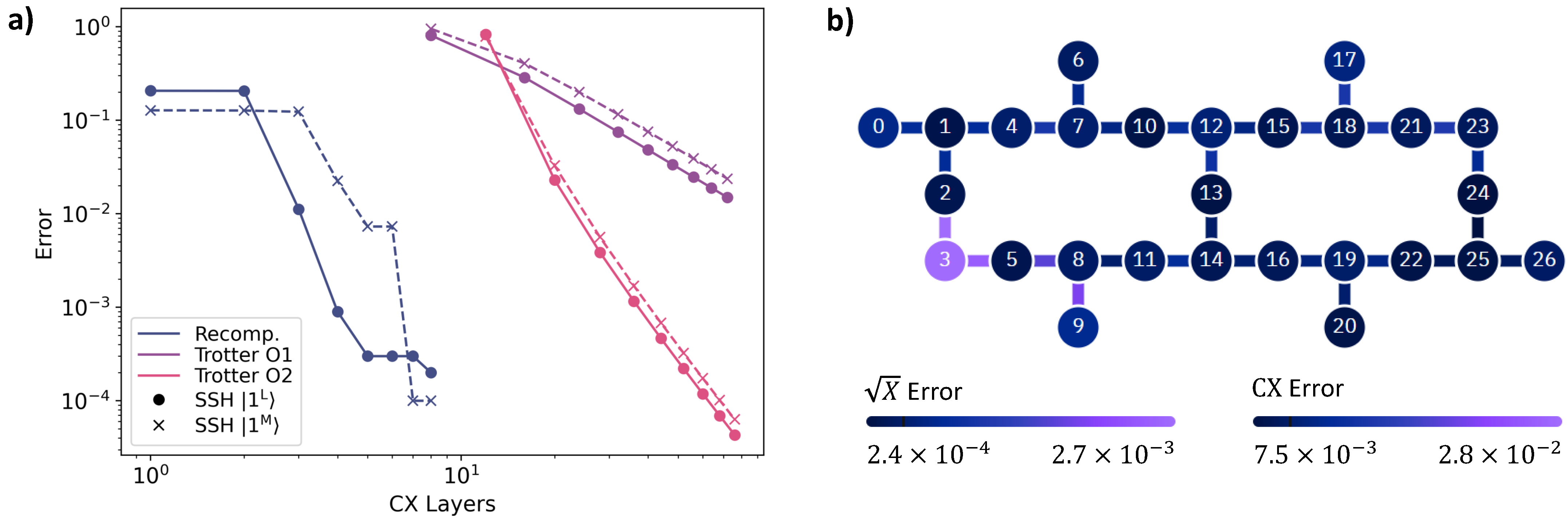}
    \caption{(a) Comparison of circuit implementation error $\mathcal{E} = 1 - \mathcal{F}(\psi^*(t), \psi(t)) = 1 - \abs{\braket{\psi^*(t)}{\psi(t)}}^2$, where $\psi^*(t) = e^{-iHt} \psi_0$ is the exact time-evolved statevector and $\psi(t)$ is that yielded by the circuit, using first- (O1) and second-order (O2) trotterization, and circuit recompilation. The $H^{\text{SSH}}$ model in the topological phase ($\nu = 1$; $w / v = 2$) is used, with initial states $\smash{\ket*{1^\text{L}}}$ and $\smash{\ket*{1^\text{M}}}$; evolution time is fixed at $t = 1$. For trotterization, the number of trotter steps ($M$) is varied; for circuit recompilation, the number of ansatz layers ($n_L$) is varied. Evidently, circuit recompilation yields far lower errors than either first or second-order trotterization, for comparable number of CX layers. (b) Illustration of qubit layout for \textit{ibmq\_paris}, with typical $\sqrt{X}$ and CX gate error rates, as reported from the IBM Q daily calibration; color shading on qubits and qubit connections reflect gate errors in correspondence to the color chart. Layout visualization and calibration data were obtained from the IBM Quantum portal~\cite{ibmq2021portal}.}
    \label{fig:trott-recomp-comp} 
\end{figure*}

We provide a comparison of the performance of circuit recompilation against first- and second-order trotterization in Figure \ref{fig:trott-recomp-comp}{a}, on the $H^{\text{SSH}}$ model. The difference in circuit depth is drastic---circuit recompilation yields comparable error but with an order of magnitude fewer CX layers. As noted, the CX cost of trotterization arises from the implementation of $\smash{e^{-i \sigma^{\vb*{\mu}} t / 2}}$ terms in each trotter step; the trotterized circuit depth for $H^{\text{KC}}$ will be significantly worse than $H^{\text{SSH}}$, due to the larger number and greater weight of the Pauli strings in the Hamiltonian. The motivation to utilize circuit recompilation is therefore abundantly clear; in fact, acquiring quantum simulation results of the same quality as that reported here is essentially infeasible, on current-generation hardware, with traditional trotterization.

\subsection{Error Mitigation}
\label{app:error}

\subsubsection{Readout Error Mitigation (RO)}
\label{app:error/ro}

Measurements on current-generation quantum hardware are imperfect---there is a chance of measuring $\ket{1}$ when a qubit is in the $\ket{0}$ state, and vice versa. A method to correct these errors is to record the measurement bit-flip probabilities on each qubit by running calibration circuits beforehand; then given raw measurement counts from experiments, the inverse problem is solved to estimate the true measurement counts~\cite{kandala2019error, kandala2017hardware, temme2017error, qiskit2020textbook, smith2019simulating}. We first discuss \textit{complete} readout mitigation. Consider a quantum circuit terminating with measurements on all $n$ qubits. Then the state of the qubits before measurement lives in Hilbert space $\mathcal{H}$ with basis states $\{\ket{s} : s \in \{0, 1\}^n\}$, that is, $s$ enumerates all binary strings of length $n$. Furthermore, suppose we have a calibration matrix $M$ with entry\footnote{Binary bitstrings are converted to their base-10 representations when acting as integer indices. This convention is followed throughout this article for notational simplicity.} $M_{ij}$ recording the probability of measuring bitstring $i \in \{0, 1\}^n$ when the true result is $j \in \{0, 1\}^n$. Denoting the raw measurement count $c_s$ for bitstring $s \in \{0, 1\}^n$, the relation between the corrected counts $c_s'$ and $c_s$ is
\begin{equation}\begin{split}
    c_s = \sum_{r \in \{0, 1\}^n} M_{rs} c_r'.
\end{split}\end{equation}

Equivalently, collating the measurement counts into $\vb{c} = [\mqty{c_0 & c_1 & \hdots & c_{2^n - 1}}]^\intercal$ and $\vb{c}' = [\mqty{c'_0 & c'_1 & \hdots & c'_{2^n - 1}}]^\intercal$, we have the linear maps $M \vb{c}' = \vb{c} \Longleftrightarrow \vb{c}' = M^+ \vb{c}$ where $M^+$ is the pseudoinverse of $M$. The estimated $\vb{c}'$ may carry negative entries due to the approximate $M$ and numerical errors; we zero these entries as they are unphysical. Note \textit{Qiskit}~\cite{qiskit2020textbook, aleksandrowicz2019qiskit} provides a procedure of least-squares fitting to estimate $\vb{c}'$ from $\vb{c}$, but we found the computational cost too large to be feasibly used for more than a handful of qubits---hence the choice of direct pseudoinverse computation. 

The matrix $M$ can be constructed by running calibration circuits~\cite{qiskit2020textbook, aleksandrowicz2019qiskit}. In circuit $\mathcal{C}_j$ we prepare $\ket{j}$ and measure all qubits; the resulting measurement probability of $i$ then sets $M_{ij}$, that is, $M_{ij} = c_i / \sum_i c_i$ from $\mathcal{C}_j$. Accordingly, complete readout mitigation requires $2^n$ calibration circuits to be run. The advantage is that correlations in measurement errors on different qubits---more precisely qubit states---are recorded; but the number of calibration circuits poses feasibility limitations. In our implementation, the required calibration circuits are prepended to each IBM Q experiment to be run. The machines available (\textit{ibmq\_manhattan}, \textit{ibmq\_paris}, \textit{ibmq\_toronto} and \textit{ibmq\_boeblingen}) limit at most $900$ circuits per job, thereby constraining $n \leq 9$ for complete readout mitigation to be usable. One may calibrate $M$ and reuse the same matrix for a period of time, so long as the same qubits are used in circuits; but the noise characteristics on hardware drift over time, and performing the calibration on-demand immediately before each experiment yields better results. 

The circuit breadth limitation motivates \textit{tensored} readout mitigation. Let us split $\mathcal{H} = \mathcal{H}_1 \otimes \mathcal{H}_2$, with basis states $\{\ket{s_1}\ket{s_2} = \ket{s_1 s_2}: s_1 \in \{0, 1\}^{n_1}, s_2 \in \{0, 1\}^{n_2}\}$. That is, we regard the quantum register of $n = n_1 + n_2$ qubits as split into two, containing $n_1$ and $n_2$ qubits respectively. Then supposing we have calibration matrices $M^{(1)}$ and $M^{(2)}$ for the two registers, we may construct $M = M^{(1)} \otimes M^{(2)}$, and $\vb{c}' = M^+ \vb{c}$ then likewise applies. In practice, it is preferable to avoid the explicit computation of $M^{(1)} \otimes M^{(2)}$ since the matrices are exponentially large; instead we write
\begin{equation}\begin{split}
    c_{s_1 s_2} = \sum_{r_1 \in \{0, 1\}^{n_1}} \sum_{r_2 \in \{0, 1\}^{n_2}} M_{r_1 s_1} M_{r_2 s_2} c_{r_1 r_2}',
\end{split}\end{equation}
and likewise compute $\vb{c}'$ using the pseudoinverses of $M^{(1)}$ and $M^{(2)}$. Note we require at most $2^{n_1} + 2^{n_2} < 2^n$ calibration circuits to obtain $M^{(1)}$ and $M^{(2)}$---in practice fewer, since each circuit calibrating the first register can be merged with one calibrating the other---but this is at the expense of neglecting correlations in measurement errors between qubits belonging to different registers. In our implementation, we choose $n_1 = \ceil{n/2}$ and $n_2 = n - n_1$, enabling feasible readout mitigation for $n \geq 13$ as needed in our experiments; for example $n = 12$ qubits requires only $96$ calibration circuits. We utilize tensored readout mitigation for all circuits containing $n \geq 10$ qubits; otherwise complete readout mitigation is used. The effect of readout mitigation on hardware data quality is smaller than that of post-selection, but it nonetheless remains an important error mitigation step; the effectiveness of post-selection diminishes when readout mitigation is not first performed.

\subsubsection{Post-Selection (PS)}
\label{app:error/ps}

Readout error mitigation reduces the effect of measurement errors, but does not mitigate errors incurred during the execution of gate rounds in the circuit. We use a post-selection procedure~\cite{mcardle2019error, smith2019simulating} to mitigate these errors to a limited extent. Suppose the Hamiltonian $H$ has symmetry (observable) $\mathcal{S}$ such that $\comm{H}{\mathcal{S}} = 0$; then $\expval{\mathcal{S}}$ is a conserved quantity. It must therefore be that $\mel{\psi(t)}{\mathcal{S}}{\psi(t)} = \mel{\psi(0)}{\mathcal{S}}{\psi(0)}$ for a time-evolved state $\ket{\psi(t)} = e^{-iHt} \ket{\psi(0)}$ and initial state $\ket{\psi(0)}$. Suppose furthermore that $\expval{\mathcal{S}}_t = \mel{\psi(t)}{\mathcal{S}}{\psi(t)}$ can be conveniently measured on the quantum circuits and $\expval{\mathcal{S}}_0$ is known; then if we find $\expval{\mathcal{S}}_t \neq \expval{\mathcal{S}}_0$, we may discard the measurement result as it is unphysical. 

The Hamiltonians studied in this article conserve particle number (see Appendix \ref{app:models}), in that $\comm{H}{O} = 0$ for total particle number operator $\smash{O = \sum_{i = 1}^{n} c^\dagger_i c_i}$ over $n$ sites. We take $\mathcal{S} = O$ for post-selection. Recall we may access site-wise occupancy expectation through computational-basis measurements on all simulation qubits (Appendix \ref{app:methods/algo/evolve}); we piggy-back on these measurements for post-selection. This leads to the following post-selection rule---suppose the raw experiment counts are $c_s$ for measurement bitstring $s \in \{0, 1\}^n$, and that the initial state $\ket{\psi(0)}$ gives known $\expval{O}_0 \in \mathbb{N}$. Then the mitigated counts are $c'_s = c_s$ if $\sum_i s_i = \expval{O}_0$ for bits $s_i \in s$, and $c'_s = 0$ otherwise. That is, we discard measurement bitstrings that have the incorrect particle number as required by symmetry.

\subsubsection{Qubit Chain Selection}
\label{app:error/chain}

The available quantum hardware provide more qubits than needed for our experiments---\textit{ibmq\_manhattan} comprises 65 qubits for instance---and there are significant variations in the quality of qubits within the same machine. It is therefore advantageous to  select qubits of the highest quality to use for experiments on an on-demand basis. Note our circuit ansatz for compilation (see Appendix \ref{app:circuit/recompilation}) requires CX connections between all adjacent pairs of qubits but not longer-range connections; hence we seek qubit chains. Given the available non-faulty qubits and CX couplings between qubits, we construct a graph representation and perform breadth-first search to identify all distinct qubit chains of the required length $n$. Then we compute the following fitness function for each chain,
\begin{equation}\begin{split}
    Q = 1 - \left\{ \prod_{i=1}^n \left( 1 - E^{U_3}_i \right)^{n_L + 1} \left[ \left( 1 - E^{\text{CX}}_{i, i-1} \right) \left( 1 - E^{\text{CX}}_{i, i+1} \right) \right]^{n_L / 2} \left( 1 - E^{\text{M}}_i \right) \right\}^{1/n},
\end{split}\end{equation}
where $E^{U_3}_i$ is the calibrated $U_3$ gate error and $E^{\text{M}}_i$ is the calibrated measurement error for qubit $i$, and $E^{\text{CX}}_{i, j}$ is the calibrated CX gate error between qubits $i$ and $j$. As before, $n_L$ is the number of layers in the circuit recompilation ansatz. The fitness $Q$ is intended to emulate the structure of the recompilation circuit ansatz; we set $n_L = 5$ as a typical value, and $E^{\text{CX}}_{1, 0} = E^{\text{CX}}_{n, n+1} = 0$ by convention. We then pick the qubit chains with highest $Q$ for experiments. Note the error rates are pulled from the daily calibration of IBM Q machines, and are hence only approximate in nature.

\subsubsection{Repetitions \& Averaging}
\label{app:error/averaging}

To minimize the effects of stochastic noise, we run each circuit for the maximum allowable number of shots $n_\text{s} = 8192$ in all our experiments. Furthermore, we perform repetitions $6 \leq n_\text{r} \leq 20$ of each circuit, such that each submitted job to the IBM Q backends saturates the $900$ circuits limit whenever possible. The effective number of shots is hence $n_\text{s} n_\text{r} \gtrsim 5 \times 10^4$ for each circuit. The degrading effects of stochastic noise on our results are further reduced by collating error-mitigated measurement bitstring counts over multiple qubit chains per machine, and multiple IBM Q machines; fluctuations in data due to noise are then averaged out. Note the performance of the machines exhibit a degree of variability---the error rates reported from the daily calibration may be inaccurate, qubits and CX connections between qubits can become faulty, and noise may cause sporadic degradation in quality of results. To filter these erroneous outliers out, we discard time series data that exhibit major discontinuities, which are unphysical; occupancy expectations must be continuous in time.

\begin{table}[!t]
    \centering
    \def\arraystretch{1.1}
    \def\padtop{-12pt}
    \def\padbottom{-8pt}
    \begin{tabular}[t]{p{0.37\textwidth} p{0.5\textwidth}}
        \begin{tabular}[t]{p{1.5cm} p{4.5cm}}
            \multicolumn{2}{l}{\textbf{SSH Model}} \\
            \toprule
            $\ket{\psi_0}$ & Initialization Circuit \\
            \midrule \\[\padtop]
                $\ket{1^{\text{L}}}$ &
                \begin{tikzcd}
                    & \gate{X} & \qw \\
                    & \qwbundle{n - 1} & \qw
                \end{tikzcd} \\ \\[\padbottom]
            \midrule \\[\padtop]
                $\ket{1^{\text{R}}}$ &
                Mirror $\ket{1^{\text{L}}}$ \\
            \midrule \\[\padtop]
                $\ket{1^{\text{M}}}$ & 
                \begin{tikzcd}
                & \qwbundle{n/2} & \qw \\
                & \gate{X} & \qw \\
                & \qwbundle{n/2 - 1} & \qw
                \end{tikzcd} \\ \\[\padbottom]
            \midrule \\[\padtop]
                $\ket{2^{\text{L}}_{11}}$ & 
                \begin{tikzcd}
                    & \gate{X} & \qw \\
                    & \gate{X} & \qw \\
                    & \qwbundle{n - 2} & \qw
                \end{tikzcd} \\ \\[\padbottom]
            \midrule \\[\padtop]
                $\ket{2^{\text{R}}_{11}}$ &
                Mirror $\ket{2^{\text{L}}_{11}}$ \\
            \midrule \\[\padtop]
                $\ket{2^{\text{L}}_{101}}$ & 
                \begin{tikzcd}
                    & \gate{X} & \qw \\
                    & \qw & \qw \\
                    & \gate{X} & \qw \\
                    & \qwbundle{n - 3} & \qw
                \end{tikzcd} \\ \\[\padbottom]
            \midrule \\[\padtop]
                $\ket{2^{\text{R}}_{101}}$ &
                Mirror $\ket{2^{\text{L}}_{101}}$ \\
            \midrule \\[\padtop]
                $\ket{2^{\text{M}}}$ & 
                \begin{tikzcd}
                & \qwbundle{n/2 - 1} & \qw \\
                & \gate{X} & \qw \\
                & \gate{X} & \qw \\
                & \qwbundle{n/2 - 1} & \qw
                \end{tikzcd} \\ \\[\padbottom]
            \midrule \\[-10pt]
                $\ket{2^{\text{LR}}}$ & 
                Concatenate $\ket{1^{\text{L}}}$ and $\ket{1^{\text{R}}}$ \\
            \bottomrule
        \end{tabular}
        &
        \begin{tabular}[t]{p{4cm} p{4.5cm}}
            \multicolumn{2}{l}{\textbf{KC Model (BDI Class)}} \\
            \toprule
            $\ket{\psi_0}$ & Initialization Circuit \\
            \midrule \\[\padtop]
                $\ket{1^{\text{L}}_j}$ for $j \in \{1, 2\}$ &
                \begin{tikzcd}
                    & \qwbundle{2(j - 1)} & \qw & \qw \\
                    & \gate{H} & \ctrl{1} & \qw \\
                    & \gate{X} & \targ{} & \qw \\
                    & \qwbundle{n - 2j} & \qw & \qw
                \end{tikzcd} \\ \\[\padbottom]
            \midrule \\[\padtop]
                $\ket{1^{\text{R}}_j}$ for $j \in \{1, 2\}$ &
                \begin{tikzcd}
                    & \qwbundle{n - 2j} & \qw &\qw & \qw \\
                    & \gate{X} & \gate{H} & \ctrl{1} & \qw \\
                    & \gate{X} & \qw      & \targ{} & \qw \\
                    & \qwbundle{2(j - 1)} & \qw & \qw & \qw
                \end{tikzcd} \\ \\[\padbottom]
            \midrule \\[\padtop]
                $\ket{1^{\text{M}}}$ & 
                \begin{tikzcd}
                    & \qwbundle{n/2 - 1} & \qw \\
                    & \gate{H} & \ctrl{1} & \qw \\
                    & \gate{X} & \targ{} & \qw \\
                    & \qwbundle{n/2 - 1} & \qw
                \end{tikzcd} \\ \\[\padbottom]
            \midrule \\[-10pt]
                $\ket{2^{\text{L}}}$ & 
                Concatenate $\ket{1^{\text{L}}_1}$ and $\ket{1^{\text{L}}_2}$ \\
            \midrule \\[-10pt]
                $\ket{2^{\text{R}}}$ & 
                Concatenate $\ket{1^{\text{R}}_1}$ and $\ket{1^{\text{R}}_2}$ \\
            \midrule \\[-10pt]
                $\ket{2^{\text{LR}}_{jk}}$ for $j, k \in \{1, 2\}$ & 
                Concatenate $\ket{1^{\text{L}}_j}$ and $\ket{1^{\text{R}}_k}$ \\
            \midrule \\[\padtop]
                $\ket{2^{\text{M}}}$ & 
                \begin{tikzcd}
                    & \qwbundle{n/2 - 2} & \qw \\
                    & \gate{H} & \ctrl{1} & \qw \\
                    & \gate{X} & \targ{} & \qw \\
                    & \gate{H} & \ctrl{1} & \qw \\
                    & \gate{X} & \targ{} & \qw \\
                    & \qwbundle{n/2 - 2} & \qw
                \end{tikzcd} \\ \\[\padbottom]
            \bottomrule
        \end{tabular}
    \end{tabular}
    \caption{Initialization circuit components for the SSH and the BDI-class KC models, corresponding to the initialization blocks drawn in the diagrams of Figure \ref{fig:circuits}. Initial states for the D-class KC model are more complicated, and are absorbed into the circuit recompilation ansatz to minimize circuit depth.}
    \label{tab:state-initialization}
\end{table}

\clearpage
\bibliography{ref-qc,ref-topo}

\end{document}